\documentclass[conference]{IEEEtran}
\usepackage[utf8]{inputenc}
\usepackage{todonotes}
\usepackage{amsmath}
\usepackage{amsthm}
\usepackage{tcolorbox}
\usepackage{amssymb}
\usepackage{floatpag}
\usepackage{textcomp} 

\usepackage[n,advantage,operators,sets,adversary,landau,notions,logic,ff,mm,primitives,events,complexity,asymptotics,keys,probability]{cryptocode}
\usepackage{wasysym}
\usepackage{wrapfig}
\usepackage{url}
\usepackage{cuted}

\usepackage{ifthen}

\definecolor{turq}{RGB}{31,115,87}

\definecolor{pink}{RGB}{207,52,118}
\newcommand{\alisachanges}[1]{\textcolor{black}{#1}}
\newcommand{\lizachanges}[1]{\textcolor{black}{#1}}
\newcommand{\peeterchanges}[1]{\textcolor{black}{#1}}

\newcommand{\gray}[1]{\textcolor{gray}{#1}}

\newcommand{\zenv}{\mathcal{Z}}
\newcommand{\ciphertxtset}{\mathsf{C}}
\newcommand{\plaintxtset}{\mathsf{M}}
\newcommand{\pkset}{\mathsf{PK}}
\newcommand{\cskset}{\mathsf{SK}_\mathrm{C}}
\newcommand{\sskset}{\mathsf{SK}_\mathrm{S}}
\newcommand{\kgenp}[1]{\mathcal{KG}_\mathrm{#1}}
\newcommand{\enca}{\mathcal{E}\mathit{nc}}
\newcommand{\encasim}{\mathcal{E}\mathit{nc}_{\simm}}
\newcommand{\decp}[1]{\mathcal{DC}_\mathrm{#1}}

\newcommand{\ver}{\mathsf{Verify}}
\newcommand{\chck}{\mathsf{Check}}

\newcommand{\oschiii}{\mathsf{DVPS}}
\newcommand{\ints}[1]{\mathbf{I}({#1})}
\newcommand{\coins}{\mathbf{Cn}}
\newcommand{\dhp}[7][{}]{\mathsf{DHP}^{#1}[{#2}\,\brokenvert\,{}_{{#3},{#4}}^{{#5},{#6}} \ifthenelse{\equal{#7}{\Box}}{}{| {#7}} ]}
\newcommand{\chp}[7][{}]{\mathsf{ChP}^{#1}[{#2}\,\brokenvert\,{}_{{#3},{#4}}^{{#5},{#6}} \ifthenelse{\equal{#7}{\Box}}{}{| {#7}} ]}
\newcommand{\pkexp}[6][{}]{\mathsf{KnE}^{#1}_{#6}[{#2}\,\brokenvert\,{}_{#3}^{#4} \ifthenelse{\equal{#5}{\Box}}{}{| {#5}}]}
\newcommand{\pkchk}[6][{}]{\mathsf{ChE}^{#1}_{#6}[{#2}\,\brokenvert\,{}_{#3}^{#4} \ifthenelse{\equal{#5}{\Box}}{}{| {#5}}]}
\newcommand{\dvp}[9][{}]{\mathsf{DVP}^{#1}({#2},{#3},{#4}\,\brokenvert\,{}_{{#5},{#6}}^{{#7},{#9}} | {#8})}
\newcommand{\dvc}[7][{}]{\mathsf{DVC}^{#1}({#2}\,\brokenvert\,{}_{{#3},{#4}}^{{#5},{#7}} | {#6})}

\newcommand{\ek}{\pckeystyle{ek}}
\renewcommand{\pcassert}{\highlightkeyword{assert}}
\newcommand{\pcget}{\highlightkeyword{get}}
\newcommand{\pcfrom}{\highlightkeyword{from}}
\newcommand{\comma}{,}
\newcommand{\equals}{=}
\newcommand{\simm}{\mathcal{S}}
\newcommand{\idfun}[1]{\mathcal{F}^{#1}}
\newcommand{\fideal}{\idfun{}}

\newcommand{\foutput}[2]{(\mathsf{Output},{{#1}},{{#2}})}
\newcommand{\fro}[2]{\mathsf{RO}({{#1}})}
\newcommand{\knetable}[1]{\mathsf{T}^{\mathsf{KNE}}_{{#1}}}

\newcommand{\symenc}[2]{\mathsf{SE}({#1},{#2})}
\newcommand{\symdec}[2]{\mathsf{SD}({#1},{#2})}

\theoremstyle{definition}

\newtheorem{theorem}{Theorem}

\begin{document}
\title{Privacy-preserving server-supported decryption}

\author{\IEEEauthorblockN{Peeter Laud}
\IEEEauthorblockA{Cybernetica AS\\
Tartu, Estonia\\
peeter.laud@cyber.ee}
\and
\IEEEauthorblockN{Alisa Pankova}
\IEEEauthorblockA{Cybernetica AS\\
Tartu, Estonia\\
alisa.pankova@cyber.ee}
\and
\IEEEauthorblockN{Jelizaveta Vakarjuk}
\IEEEauthorblockA{Cybernetica AS\\
Tallinn University of Technology\\
Tallinn, Estonia\\
jelizaveta.vakarjuk@cyber.ee}}

\maketitle

\begin{abstract}
In this paper, we consider encryption systems with two-out-of-two threshold decryption, where one of the parties (the client) initiates the decryption and the other one (the server) assists. Existing threshold decryption schemes disclose to the server the ciphertext that is being decrypted. We give a construction, where the identity of the ciphertext is not leaked to the server, and the client's privacy is thus preserved. While showing the security of this construction, we run into the issue of defining the security of a scheme with blindly assisted decryption. We discuss previously proposed security definitions for similar cryptographic functionalities and argue why they do not capture the expected meaning of security. We propose an ideal functionality for the encryption with server-supported blind threshold decryption in the universal composability model, carefully balancing between the meaning of privacy, and the ability to implement it. We construct a protocol and show that it is a secure implementation of the proposed functionality in the random oracle model.
\end{abstract}

\pagestyle{plain} 
\thispagestyle{plain}
\section{Introduction}\label{sec:intro}

Threshold cryptography offers a set of techniques for management of private keys; it can help out if there is no single party sufficiently trusted and sufficiently in control of its computational environment, such that they could be allowed to know the whole private key. Both threshold signatures~\cite{threshSTOC94} and threshold decryption~\cite{shoupgennaro} are well-known cryptographic techniques. For signatures, there exist large-scale deployments in blockchain environments~\cite{DBLP:conf/ccs/LindellN18, doerner2019threshold}, as well as for general electronic identity~\cite{splitkey}. In the last example, threshold cryptography is used because a certain party does not have sufficient control over their computational environment. That party is a \emph{user}, having a \emph{smartphone} that stores their keyshare. If one wants to avoid secure hardware based solutions~\cite{secureelement}, the phone alone cannot offer sufficient protection for a private key.

We emphasize that in the use-cases related to the electronic identity, we are specifically interested in 2-out-of-2 secret sharing of the private key. Sharing is necessary, because otherwise the server alone would be able to use the private key. Having just two parties is natural in such system --- the roles of the phone and the assisting server are clearly different, while any more parties (e.g. implementing the server in threshold manner) would reduce the usability of the system, would not significantly increase the security (the server is expected to securely manage the keyshare), and may even interfere with some of the mechanisms for detecting break-ins.

\peeterchanges{In the near future, a user's digital presence is constructed around the credentials that they have obtained from the various issuers and are able to present to the various relying parties (RP). There exist standards, e.g. for mobile Driver's License~\cite{mdl} by ISO or for verifiable credentials~\cite{w3ccreds} by W3C that specify, how these credentials are formatted and through which protocols they are presented. There exist advanced legistlative initiatives, e.g. the upcoming eIDAS~2.0 regulation of the European Union~\cite{eidas2,eidas20finalagreement} that will mandate the availability of \emph{digital identity wallets} --- smartphone applications that manage the user's credentials by following these standards, and implementing a certain architecture~\cite{eudiw}. The architecture may, to a certain extent, protect the user's privacy through \emph{selective disclosure}, releasing only user-approved parts of multi-part credentials to a RP~\cite{mdl,sd-jwt}, but more capable privacy techniques, e.g. zero-knowledge proofs~\cite{zkp} are not part of the architecture.}

\peeterchanges{The wallet app running on the smartphone has to manage the storage of credentials. They may be stored on- or off-device, but will definitely be encrypted, because their content may be sensitive. Before a presentation of the credential, it has to be decrypted, because the wallet app needs to know its content to be presented. The app and the phone have to have control over the decryption step.} If one wants to avoid secure hardware based solutions, then the decryption has to be thresholdized, and the same 2-out-of-2 setting makes the most sense. In this case, whenever an RP asks the user to present a credential that the user has and the user agrees to present it, the phone sends the ciphertext encrypting that credential to the assisting server. The phone and the assisting server run the threshold decryption protocol, and (only) the phone learns the plaintext credential. After running the protocol(s) with the RP, the phone deletes the decrypted credential from its memory. Next time the presentation of a credential is requested, the phone and the assisting server again run the decryption protocol. Any threshold decryption protocol is in principle usable here, e.g. the original ElGamal-based construction of Shoup and Gennaro~\cite{shoupgennaro}.

Unfortunately, with regular threshold decryption protocol, the server will learn how often one or another ciphertext is being decrypted. This is a potential violation of the user's privacy: if, over time, the server sees several requests to decrypt the same credential, then it may make some inferences on the user's access pattern towards different RPs. This paper proposes a solution for this problem: we construct a 2-out-of-2 threshold encryption scheme, where the decryption protocol is privacy-preserving against the server. Aiming for generality, our scheme is secure against chosen-ciphertext attacks even when one of the decrypting parties (either the client or the server) is corrupted.

Definitions of security properties for such schemes are similar to those for the encryption schemes with \emph{blind assisted decryption}. Unfortunately, we have found existing definitions to not correspond to our intuitive notions of security; the discussion of these definitions is the second main contribution of this paper. We follow that discussion by proposing our own security definition for the server-assisted blind decryption. While the previous definitions have been given in the game-based model, stating the security game(s) that an adversary attempts to win, our definition is given in the universal composability (UC)~\cite{UC} framework \alisachanges{where the desired security properties are covered by an ideal functionality describing how the protocol \emph{should} work. As there are several distinct properties that we are aiming for, we would need several different game-based definitions, which make it more complicated to justify why exactly these properties are necessary and sufficient, while an ideal functionality provides a generic overview.} In this way, we get a better understanding of the details of the security definition, hopefully giving us a better assurance that it is the right one. Still, we see that our definitions follow a fine line between the intuitive meaning of the security and privacy on one side, and the implementability on the other side; we explore some new ground in defining UC encryption, and the details become significant. \alisachanges{We see that the blinding makes analysis of composability quite important in the malicious client case. Due to the blinding, the client can potentially combine several server responses (which are not necessarily successful decryption outputs) in such a way that the plaintext (or even several plaintexts at once) will not be learned until the last decryption, and some attacks by a malicious client (in particular, the ability to decrypt more ciphertexts than there have been sessions with the server initiated) could be missed in the stand-alone model.} 

In the construction of our threshold encryption scheme, the goal has been to show the feasibility, and the conceptual simplicity of presentation. Still, its performance is fully satisfactory for the use-case of verifiable credentials. It should also be simple to implement and deploy, as it uses only simple cryptographic primitives, in particular avoiding pairings.

This paper has the following structure. After reviewing related work in Sec.~\ref{sec:relwork}, we will dive into the previous security definitions of blind decryption in Sec.~\ref{sec:definitions}. Next, we describe our security definition by presenting corresponding ideal functionality in Sec.~\ref{sec:idfun}. In Sec.~\ref{sec:ourconstr}, we present main building blocks of our protocols and give our construction of a privacy-preserving threshold encryption scheme. Finally, we prove in Sec.~\ref{sec:proof} that our construction indeed satisfies our security definition.

\section{Related work}\label{sec:relwork}
Encryption schemes with threshold decryption~\cite{shoupgennaro} and with indistinguishability against chosen-ciphertext attacks (IND-CCA) were proposed shortly after IND-CCA secure asymmetric encryption schemes~\cite{cramershoup}. At present, threshold cryptography is a mature field, discussed in textbooks~\cite{bonehshoupbook} and subject to standardization activities~\cite{NISTIR8214}. However, we note the dearth of the exploration of threshold decryption in the UC model, meaning that many details of our ideal functionality are novel.

Regarding the ideal functionalities for the usual public key encryption, several different approaches can be taken. One can design the ideal functionality so, that the adversary internally executes the encryption and decryption algorithms~\cite{cryptoeprint:2003/174}. Alternatively, the adversary may give descriptions of encryption and decryption algorithms to the ideal functionality, with the latter invoking them locally~\cite{UC-2005}. In the former case, the adversary may cause malicious ciphertexts to decrypt to values that depend on the values to which legitimate ciphertexts are decrypted to~\cite{UC-2005}.

Our target use case involves two non-equal parties. In this setting, Buldas et al.~\cite{splitkey} have proposed a 2-out-of-2 sharing for \emph{signing} keys, where one of the shares is stored in a central server and the other one in the user's smartphone, encrypted with only a PIN. The signing protocol employs measures against an adversary guessing that PIN, or making a copy of the memory of the smartphone. Lueks et al.~\cite{tandem} have proposed a system with the same kind of 2-out-of-2 sharing, where also the users' usage patterns are protected; the system also relies on blind signatures. For the threshold decryption, Kirss et al.~\cite{kirss-et-al} have proposed a protocol with the same measures for protecting the keyshare in the smartphone as~\cite{splitkey}; together with the clone detection mechanisms of~\cite{Sarr19}, it could offer a viable alternative to secure elements~\cite{secureelement}. Unfortunately, they do not attempt to hide the ciphertext from the assisting server.

Our scheme offers privacy for one of the decrypting parties against the other decrypting party. In this sense, our scheme is an instance of \emph{blind assisted decryption}. We are aware of two previous attempts to formalize the IND-CCA security of blind assisted decryption, and devise schemes satisfying these definitions. Green~\cite{green} proposed a scheme, where the client sends decryption requests to a server with the private key. Blazy et al.~\cite{blazy} proposed an encryption system where the decryption capability and authorization was shared among three parties with unequal roles, offering privacy against the ``server'' party. We discuss both schemes and corresponding security definitions in Sec.~\ref{sec:definitions}.

We analyse the security of our scheme in the universal composability model with ROM~\cite{UC}. We use a less common hardness assumption: the \emph{one-more static computational Diffie-Hellman problem}~\cite{DBLP:conf/pkc/Boldyreva03,one-more}. Still, the assumption is not controversial: it is known to hold in the Generic Group Model and be equivalent to the hardness of Discrete Logarithm in the Algebraic Group Model~\cite{cryptoeprint:2021/866}. The one-more problems commonly occur in the security analysis of \emph{blind signatures}~\cite{DBLP:conf/pkc/Boldyreva03,DBLP:journals/joc/BellareNPS03}, hence one may even expect it to make an appearance for our blind primitive as well.

\section{Security of blind assisted decryption}\label{sec:definitions}

Blind assisted decryption has been previously considered by Green~\cite{green} and by Blazy et al.~\cite{blazy}. Both works have separately defined the blindness property, and the property of being secure against chosen ciphertext attacks. While the first property is not controversial, the proposed definitions for the latter cannot be considered fully satisfactory.

Blazy et al. consider a setting with three parties --- \emph{user}, \emph{token}, and \emph{server}. The token and the server have keyshares $\mathit{sk}_\mathcal{T}$ and $\mathit{sk}_\mathcal{S}$ of a 2-out-of-2 threshold decryption key, while the user and the server share a password $\mathit{pwd}$. The user initiates the decryption by activating the token with the ciphertext. The token then runs a protocol with the server. They define a password-protected Indistinguishability under Replayable Chosen-Ciphertext Attacks (P-IND-RCCA). Their multi-party (i.e. there are several users, tokens, and servers, each of those may be corrupted by the adversary) definition provides the adversary with a decryption oracle that performs decryption atomically (unless the submitted ciphertext decrypts to one of the two challenge plaintexts, as is the norm for RCCA~\cite{DBLP:conf/crypto/CanettiKN03}). When submitting these challenge plaintexts, the adversary specifies the identities of the user $\mathcal{U}^*$, token $\mathcal{T}^*$, and server $\mathcal{S}^*$ that it is attacking. The adversary also has access to the message-sending oracle, where it submits a message to one of the parties, and learns the message that this party replies with.

In order to model that no protection is offered if the adversary corrupts too many parties, the definition of P-IND-RCCA restricts, which parties the adversary may invoke. If $\mathcal{U}^*$ or $\mathcal{S}^*$ are corrupt (i.e. the adversary knows the password), then $\mathcal{T}^*$ may no longer be invoked. If $\mathcal{U}^*$ and $\mathcal{T}^*$ are corrupt (i.e. the adversary has the password and one of the keyshares) then $\mathcal{S}^*$ may no longer be invoked. Indeed, in such cases, ability to make queries to these parties might allow the adversary to decrypt the challenge ciphertext.

Such restrictions make P-IND-RCCA more akin to the ``lunchtime'' (CCA1) attack, when considering message-sending oracle. Also, some intuitively insecure encryption schemes satisfy P-IND-RCCA. Consider any P-IND-RCCA secure scheme, and modify it as follows:
\begin{enumerate}
    \item Whenever the server $\mathcal{S}$ receives a message, it will check whether it has the format $(\mathsf{corrupt}, \mathit{pwd}_\mathcal{U},\mathit{sk}_\mathcal{T})$, where $\mathcal{U}$ and $\mathcal{T}$ are the user and the token in the current decryption session, $\mathit{pwd}_\mathcal{U}$ is user's password, and $\mathit{sk}_\mathcal{T}$ is the keyshare of the token. The check can be done by encrypting a random plaintext and then running the decryption protocol in server's head. If the incoming message has such format, then respond with its own keyshare $\mathit{sk}_\mathcal{S}$.
    \item \emph{Alternatively}, when the token $\mathcal{T}$ receives the message $(\mathsf{corrupt},\mathit{sk}_\mathcal{S})$, the ``correctness'' of which it can check in the same manner, then it responds with $\mathit{sk}_\mathcal{T}$.
\end{enumerate}
We argue that the modified scheme (with either the first or the second change) should not be considered secure any more, because an adversary that has managed to take over one of the parties holding a secret share can easily extend this takeover to obtaining the other share. But the modified scheme still satisfies P-IND-RCCA, because such messages cannot be sent according to this definition. Indeed, an adversary that has corrupted the user and the token (giving him $\mathit{pwd}_\mathcal{U}$ and $\mathit{sk}_\mathcal{T}$) in the session being attacked, may no longer talk to the server. Similarly, after corrupting the server and learning $\mathit{sk}_\mathcal{S}$, the adversary may no longer talk to the token.

Green~\cite{green} considers a user $\mathcal{U}$ outsourcing decryption to a server $\mathcal{D}$. The server has the decryption key; the user, having a ciphertext, initiates the blind decryption protocol. The security of their scheme against chosen ciphertext attacks is captured by a ``usual'' IND-CCA2 definition, where the adversary can atomically invoke the ``normal'' decryption oracle, and by the \emph{leak-freeness} definition. In this definition, a Distinguisher is asked to distinguish between a ``real'' and an ``ideal'' game. In the real game, an adversary chooses a ciphertext and the Distinguisher gets the user's view in the blind decryption protocol. Note that the execution of the blind decryption protocol is \emph{atomic}, i.e. the adversary does not interfere with its run. In the ideal game, a simulator chooses a ciphertext, receives the corresponding plaintext, and simulates the trace. I.e. leak-freeness ``ensures that an adversarial user gains no more information from the blind decryption protocol than they would from access to a standard decryption oracle''~\cite{green}.

The absence of the blind decryption protocol in the IND-CCA2 definition and/or the atomicity requirement in the set-up of the ``real'' game allows some intuitively insecure schemes to satisfy Green's IND-CCA2 and leak-freeness definitions. Consider any scheme that satisfies them, and modify its blind decryption as follows:
\begin{itemize}
    \item All messages from $\mathcal{U}$ to $\mathcal{D}$ have an additional bit $b_\mathsf{evil}$.
    \item If $\mathcal{D}$ receives a message with $b_\mathsf{evil}=0$, then it processes the rest of the message as in the original scheme.
    \item If $\mathcal{D}$ receives a message with $b_\mathsf{evil}=1$, then it responds with the decryption key.
    \item $\mathcal{U}$ always sets $b_\mathsf{evil}=0$.
\end{itemize}
Such scheme should definitely be considered as insecure. But it satisfies both IND-CCA2 (because blind decryption protocol is not invoked there) and leak-freeness (because the atomic execution of the blind decryption protocol means that no one sets $b_\mathsf{evil}=1$).

\section{Ideal functionality for privacy-preserving server-assisted decryption}\label{sec:idfun}

We consider asymmetric encryption schemes, where the decryption functionality is distributed between two parties --- the \emph{client}, and the \emph{server} with different roles. 
An encryption scheme with the client-server decryption consists of the following sets, algorithms, and protocols, all parameterized with the security parameter $\lambda$. 

\begin{itemize}
\item Sets of ciphertexts $\ciphertxtset$, public keys $\pkset$, the client's private keys $\cskset$, and the server's private keys $\sskset$.
\item Key-generation protocol $\langle\kgenp{C}|\kgenp{S}\rangle$, run by both parties. It returns $(\sk_C,\pk)\in\cskset\times\pkset$ to the client, and $(\sk_S,\pk)\in\sskset\times\pkset$ to the server.
\item Encryption algorithm $\enca$. It takes as input a public key $\pk\in\pkset$, a plaintext $m\in\plaintxtset$ and returns a ciphertext $c\in\ciphertxtset$.
\item Decryption protocol $\langle\decp{C}|\decp{S}\rangle$, run by the client and the server. Client's inputs to $\decp{C}$ are $c\in\ciphertxtset$, $\sk_C\in\cskset$, and $\pk\in\pkset$. Server's inputs to $\decp{S}$ are $\sk_S\in\sskset$ and $\pk\in\pkset$. The protocol returns either $m\in\plaintxtset$ or the failure notice $\bot$ to the client. It returns the success notice $\top$ or the failure notice $\bot$ to the server.
\end{itemize}

\alisachanges{First of all, we define how such a scheme \emph{should} work. Both functional and security requirements can be covered by an \emph{ideal functionality} in the Universal Composability framework (UC)~\cite{UC}.} \lizachanges{This framework has been widely used to capture security of distributed protocols. It allows to capture security properties of cryptographic schemes in the ideal/real process paradigm. In the real world, adversary $\adv$ interacts with the real protocol $\pi$ \alisachanges{by means of corrupting different parties, gaining control over their inputs and outputs}. In the ideal world, ideal functionality \alisachanges{$\fideal$} defines an ideal process for the protocol $\pi$, \alisachanges{intuitively describing how it should work}. Ideal functionality \alisachanges{can be viewed as} a trusted party that receives inputs from all the parties, performs computation on these inputs, and returns to each party its output. \alisachanges{It also interacts with an ideal adversary $\simm$, from which it may receive certain commands and to which it may output some values that are explicitly allowed to be leaked.} In both cases, parties receive inputs and return outputs to the environment $\zenv$. A protocol $\pi$ securely implements $\idfun{}$, if for any real-world adversary $\adv$ there exists an ideal adversary $\sdv$, such that no environment $\zenv$ can distinguish between interaction with $\pi$ \alisachanges{running in parallel with} $\adv$ and $\idfun{}$ \alisachanges{running in parallel with} $\sdv$.}

The ideal functionality for our encryption scheme with two-party privacy-preserving decryption is given in Fig.~\ref{fideal}. For simplicity, we define it only for a single public key (and the corresponding shared decryption functionality). Also for simplicity, we consider only static corruptions of parties; the identities of the corrupt parties are told to the $\fideal$ by the ideal adversary $\simm$ at the beginning of the execution. The functionality $\fideal$ is given for an arbitrary number of parties $P_1,P_2,\ldots$, all of which can submit encryption requests. The decryption can be done jointly by parties $P_1$ (in the role of the client) and $P_2$ (in the role of the server).

\begin{figure*}[tbp]
\begin{tcolorbox}[colback=white,arc=0.3mm, boxrule=0.3mm,fontupper=\normalsize]

\underline{Key generation} \\
On message $(\textsf{KeyGen}, sid)$ from both $P_1$ and $P_2$:
\begin{enumerate}
    \item Send $(\textsf{KeyGen},sid)$ to $\simm$ and wait for $(\textsf{Key},sid, \pk)$ from $\simm$.
    \item \alisachanges{If $\pk \neq \bot$,} create an empty table $\mathsf{T}$ for storing encrypted messages, and set decryption counter $\mathit{ctr}_\mathsf{dec}\gets 0$.
    \item\label{fideal:keygen:return} \alisachanges{For $i \in \{1,2\}$ , if the party $P_i$ is corrupted, wait for $\foutput{i}{y_i}$ from $\adv$. Otherwise, take $y_i = \pk$.
    \item Output $(\textsf{Key}, sid, y_1)$ to $P_1$ and $(\textsf{Key}, sid, y_2)$ to $P_2$.}
\end{enumerate}
\vspace{0.2cm}

\underline{Encryption} \\
On message $(\textsf{Encrypt},sid,m)$ from some party $P_i$:
\begin{enumerate}

    \item If $P_i$ is corrupted, send $(\textsf{Encrypt},sid,\alisachanges{i},m)$ to $\simm$. Otherwise, send $(\textsf{Encrypt},sid)$ to $\simm$.
    \item Wait for $(\textsf{Encrypt},sid,c)$ from $\simm$ (the adversary chooses the ciphertext for the given plaintext).
    \item Store the pair $(m,c)$ in table $\mathsf{T}$.
    \item\label{fideal:encrypt:return} \alisachanges{If the party $P_i$ is corrupted, wait for $\foutput{i}{y}$ from $\adv$. Otherwise, take $y = c$.}
    \item Send $(\textsf{Encrypted},sid,\alisachanges{y})$ to $P_i$.

\end{enumerate}
\vspace{0.2cm}

\underline{Decryption procedure} $Decrypt(\dec, c)$ for a function $\dec : \ciphertxtset \to \plaintxtset$ and a ciphertext $c \in \ciphertxtset$:
        \begin{itemize}
        \item If there exists a pair $(m,c)$ in $\mathsf{T}$: (for correctness we need that $c$ would decrypt to $m$)
        \begin{itemize}
            \item If $m$ is not unique (there are two different $m$ for the same $c$), take $m' = \bot$.
            \item If $m$ is unique, take $m' = m$.
        \end{itemize}
        \item Otherwise, compute $m' := \dec(c)$, running it only a polynomial number of steps (the polynomial is a parameter of $\fideal$). Add $(m',c)$ to $\mathsf{T}$.
        \item Return $m'$.
        \end{itemize}
\vspace{0.2cm}

\underline{Decryption} (honest client) \\
On message $(\textsf{Decrypt},sid,c)$ from party $P_1$ and $(\textsf{Decrypt},sid)$ from party $P_2$:
\begin{enumerate}
\item Send $(\textsf{Decrypt-init}, sid)$ to $\simm$.
\item Upon receiving $(\textsf{Decrypt-init}, sid, \ver)$ from $\simm$, where $\ver : \ciphertxtset \to \set{\mathsf{true},\mathsf{false}}$, compute $b \gets \ver(c)$.
    \begin{itemize}
        \item If $b = \alisachanges{\textsf{true}}$, send $(\textsf{Decrypt-good-c}, sid)$ to $\simm$.
        \item If $b = \alisachanges{\textsf{false}}$ \alisachanges{or $b = \bot$}, send $(\textsf{Decrypt-bad-c}, sid)$ to $\simm$. \alisachanges{Proceed to the point~\ref{fideal:decrypt-hc:return}) with $y_1 = y_2 = \bot$.}
    \end{itemize}
\item The final output $y_1$ for $P_1$ depends on whether the server is corrupt.

    If the server is \emph{honest}, the decryption can only be delayed:
    \begin{itemize}
        \item Upon receiving $(\textsf{Decrypt-complete}, sid, \dec)$ from $\simm$, take $y_1 \gets Decrypt(\dec,c)$, and $y_2 \gets \top$.
    \end{itemize}

    If the server is \emph{corrupted}, $\simm$ tells whether decryption succeeded. 
    \begin{itemize}
        \item Upon receiving $(\textsf{Decrypt-complete}, sid, \dec)$ from $\simm$, take $y_1 \gets Decrypt(\dec,c)$. 
        \item Upon receiving $(\textsf{Decrypt-fail}, sid)$ from $\simm$, take $y_1 \gets \bot$. 
    \end{itemize}
    \item\label{fideal:decrypt-hc:return} \alisachanges{If the server is corrupted, wait for $\foutput{2}{y'_2}$ from $\adv$ and take $y_2 \gets y'_2$.}
        
\item Send $(\textsf{Decrypted},sid,y_1)$ to $P_1$ and $(\textsf{Decrypted},sid,y_2)$ to $P_2$.
\end{enumerate}
\vspace{0.2cm}

\underline{Decryption} (corrupted client) \\
On message $(\textsf{Decrypt},sid,c)$ from party $P_1$ and  $(\textsf{Decrypt},sid)$ from party $P_2$:

\begin{enumerate}
    \item Set $\mathit{ctr}_\mathsf{dec} \gets \mathit{ctr}_\mathsf{dec}+1$ and send $(\textsf{Decrypt-init}, sid, c)$ to $\simm$.
    \item $\simm$ tells whether decryption succeeded:
    \begin{itemize}
        \item Upon receiving $(\textsf{Decrypt-complete}, sid)$ from $\simm$, take $y_2 \gets \top$.
        \item Upon receiving $(\textsf{Decrypt-fail}, sid)$ from $\simm$, take $y_2 \gets \bot$.
\end{itemize}    
    \item\label{fideal:decrypt-cc:return} \alisachanges{Wait for $\foutput{1}{y_1}$ from $\adv$.}
\item Send $(\textsf{Decrypted},sid,y_1)$ to $P_1$ and $(\textsf{Decrypted},sid,y_2)$ to $P_2$.
\end{enumerate}

\underline{At any point of time \alisachanges{(corrupted client)}.} On message $(\textsf{Decrypt-msg}, sid, \dec, c')$ from $\simm$:
\begin{enumerate}
    \item Compute $m = Decrypt(\dec,c')$.
    \item Set $\mathit{ctr}_\mathsf{dec}\gets\mathit{ctr}_\mathsf{dec} - 1$. If $\mathit{ctr}_\mathsf{dec}<0$, then stop.
    \item Send $(\textsf{Decrypted}, sid, m)$ to $\simm$.
\end{enumerate}

\end{tcolorbox}
\caption{Ideal encryption functionality $\fideal$}.\label{fideal}
\end{figure*}

The key generation (that only happens in the beginning) and encryption in $\fideal$ are similar to existing ideal functionalities for public-key encryption, where the public key $\pk$ is chosen by $\simm$. 
Encryption is performed similarly to~\cite{cryptoeprint:2003/174}, where the adversary comes up with the ciphertext $c$ for the plaintext $m$ without actually seeing $m$, intuitively ensuring that $c$ does not leak anything about $m$. Only if the encryptor party $P_i$ is corrupted, will $\simm$ learn the message $m$. The obtained pairs $(m,c)$ are stored in a table $\mathsf{T}$, allowing the decryption queries to be answered. 

The encryption query does not contain a public key as an input. In practice, the environment could provide an honest encryptor with a public key $\pk'$ different from $\pk$ generated by $\fideal$, and expect a ciphertext encrypted with $\pk'$. To cover this case, we would need to allow $\fideal$ deliver $\pk'$ to $\simm$. This would introduce more details into the definition of $\fideal$ and the security proofs without being essential for the discussion of the key points of $\fideal$ and its secure implementations. We assume that a party (in the environment) only makes encryption queries once he has learned the correct $\pk$.

Decryption in $\fideal$ corresponds to finding the plaintext $m$ from a ciphertext $c$ that was created using the key $\pk$. It can only be initiated by $P_1$ and $P_2$, with $P_1$ providing the ciphertext.

If the client is honest, then the decryption is less straightforward than in the UC public key encryption, because neither of the two existing approaches appear to be fully satisfactory. If $\simm$ provides the descriptions of the algorithms $\enc$ and $\dec$ to $\fideal$, then it may be difficult to build simulators in the Random Oracle Model (ROM), because the simulator probably needs to know the state of the oracles during decryption, as well as program the oracles at least during encryption. But if we task $\simm$ to come up with the ciphertext values during encryption, and to decrypt unknown ciphertexts during decryption, then this goes against our intuition of ``blinded decryption'', where $\simm$ should not learn the ciphertexts that an honest client wants to decrypt. We resolve this by making $\simm$ run the encryption (programming the random oracles as necessary), and sending to $\fideal$ an \emph{updated} description of $\dec$ at each decryption. In this way, $\simm$ does not learn the ciphertext that is being decrypted, but the current state of the random oracles can be included in the description of $\dec$ and affect the result of decryption. Using this approach (and similarly to UC public key encryption), we need to be careful that the same ciphertext $c$ will not decrypt to different values, which would violate correctness. For that reason, $\fideal$ adds to $\mathsf{T}$ all $(m,c)$ pairs obtained during the decryption as well. If there will be two records $(m,c)$ and $(m',c)$ for $m \neq m'$, the decryption fails. In addition, since $\simm$ cannot verify the correctness of the input ciphertext, $\simm$ should first of all send to $\fideal$ a description of a function $\ver$ which verifies the ciphertext, returning either $\mathsf{true}$ or $\mathsf{false}$. $\simm$ needs to learn this single bit of information, as in a real protocol the decryption would not start in this case.

If the client is corrupted (but the server is honest), then privacy of the ciphertext does not have to be protected; ciphertext can be given to $\simm$. But we are now modelling threshold decryption, where a corrupted party is expected to learn the plaintext; we are not aware of any similar models in the literature. If a corrupted party is able to learn the plaintext $m$, then a simulator must also be able to learn it, in order to simulate that party's view. But if the ciphertext $c$ was created using the encryption functionality of $\fideal$, then $c$ is actually independent of $m$, and the only location containing $m$ (beside the environment, which the simulator cannot depend on) is in the table $\mathsf{T}$ of $\fideal$. Hence we have to give $\simm$ an ability, however minimal, to read that table. In the case of threshold decryption \emph{without} blinding, it would be sufficient if $\simm$ could query for the row $(m,c)$, where $c$ is the ciphertext that is being decrypted. For the blinded decryption, the adversary playing the corrupted client can actually start the decryption protocol for an arbitrary ciphertext, hence the row(s) queried by $\simm$ can no longer be restricted like that. We should allow $\simm$ to query for the row $(m,c)$ for a ciphertext $c$ its own choice. Moreover, we allow to make this query \emph{after} the decryption session has ended, since the corrupted client may potentially be able to undo his own blinding in multiple ways after getting response from the server, thus being able to choose between decrypting several different ciphertexts later. Nevertheless, the adversary is only allowed to decrypt \emph{at most one message per decryption query} coming from the environment.

We deliberately have omitted the case where both the server and the client are corrupted, as we do not aim to achieve any security guarantees in this case. The adversary would get full control over the key generation, encryption and decrypton, and would be given all messages and ciphertexts that any party (including the honest ones) receives as an input. 

Next, we discuss how the ideal functionality that we defined sidesteps the problems we identified with the previous security definitions. As a counterexample for P-IND-RCCA, we proposed a scheme where the adversary may obtain the secret key share of an honest party if he gets the secret key share of a corrupted party. If such a protocol were run in the UC model with either the client or the server corrupted, the adversary would be able to reconstruct the private key $\sk$ and decrypt the ciphertexts that have been encrypted with $\pk$. The environment could distinguish the real protocol from $\fideal$ by decrypting (without involving $\fideal$) an encryption of a message $m$ generated by $\fideal$, i.e. without letting $\fideal$ tell $m$ to $\simm$, getting a plaintext that does not depend on $m$.

Compared to the definition of Green, our execution of blinded decryption in $\fideal$ is not atomic, and the adversary has full control over the corrupted client, and thus could set up $b_{evil} = 1$. The environment would get $\sk$ and hence could distinguish the real protocol from $\fideal$ by decrypting (without involving $\fideal$) an encryption of a message $m$ generated by $\fideal$, getting a plaintext that does not depend on $m$. Non-atomic execution of blinded decryption allows a corrupted client can decrypt a ciphertext of his own choice, which would not be possible using atomic execution. We explicitly let the corrupted client access a selected entry from the table $\textsf{T}$, which can be viewed as access to a standard decryption oracle. If the client is not corrupted, then the access to the decryption oracle is limited and controlled by the environment. 

\section{Secure implementation}\label{sec:ourconstr}

\peeterchanges{While we have critisized the \emph{definitions} of Green~\cite{green} and Blazy et al.~\cite{blazy}, we may wonder whether their implementations, even though they are for different functionalities, are adaptable into a secure implementation of $\fideal$. We can immediately answer that question in negative for Green: his outsourced blinded decryption is not an instance of threshold decryption, and does not support the user independently finding out whether a ciphertext is valid. The latter is necessary for chosen-ciphertext security in threshold decryption schemes.
}

\peeterchanges{Blazy et al.~\cite{blazy} discuss whether their construction could be realizable in universally composable manner, and answer that question in negative. But that answer stems from the inability to simulate the adversary guessing the password of an honest party. It is possible that if the roles of \emph{user} and \emph{token} in their protocol were combined into one, then their construction would provide a secure implementation of $\fideal$.}

\peeterchanges{However, the blind decryption step of Blazy et al.~\cite{blazy} uses Groth-Sahai proofs~\cite{grothsahai} to prove that a blinded ciphertext has been constructed from a ciphertext that was accompanied by proofs of validity. This means that their construction is complex, and heavily based on bilinear pairings, leading to both relatively heavy computations and to large message sizes.
}

We securely implement $\fideal$ under common cryptographic assumptions: hardness of discrete logarithm (DL), one-more CDH, and the existence of additively homomorphic encryption systems, for which the proofs of equality of plaintexts can be given. In this paper, we will instantiate the latter with Paillier encryption, which is IND-CPA (indistinguishability against chosen-plaintext attacks) secure under the Decisional Composite Residuosity Assumption (DCRA)~\cite{Paillier}. Remarkably, our construction $\oschiii$ does not require bilinear pairings. \peeterchanges{Compared to~\cite{green,blazy}, a potential theoretical shortcoming of our construction is its use of the Random Oracle Model (ROM), but considering its prevalence in constructions employed in practice~\cite{rsapss,schnorr}, we do not see it as a weakness.}

Our construction $\oschiii$ builds on top of the IND-CCA secure TDH1 cryptosystem of Gennaro and Shoup~\cite{shoupgennaro}, adding to it ciphertext blinding and unblinding operations that are used when the client queries the server during decryption protocol. TDH1 builds upon the ElGamal key encapsulation mechanism (KEM), adding to its ciphertexts $c$ the non-interactive zero-knowledge (NIZK) proofs $\pi$ that someone (e.g. the entity encrypting the plaintext $m$) knows the randomness $r$ used for the encryption. In $\oschiii$, both $c$ and the proof have to be blinded. To blind the proof, we need \emph{malleable} NIZK proofs~\cite{malleableZK}, but the known constructions are based on bilinear pairings. Fortunately, we do not need malleable NIZK in its full generality \peeterchanges{(as used by Blazy et al.~\cite{blazy})}; we only need the server to be convinced by a proof that the client malleated. There exist malleable \emph{designated verifier proofs} without pairings~\cite{DFN} that will be used in our construction.

Our ciphertexts will thus contain a proofs of knowledge of the randomness $r$, designated to be verified by the server. The client also needs to be convinced that $r$ is known; we use ``usual'' NIZK proofs for that. Our ciphertexts also contain a third kind of proof, convincing the client that an honest server is going to accept the designated verifier proof.
\alisachanges{In~\cite{cryptoeprint:2017/1029}, the DFN proofs of~\cite{DFN} have been found vulnerable against selective failure attacks that allow to find the secret of the verifier bit by bit. This can be mitigated by letting the server stop after a certain number of failed proofs and require a fresh key generation. The ability of the client to verify whether the server will accept the proof is needed to avoid attacks where an external encrypting party produces bad ciphertexts that will be rejected by the server and cause denial of service for an honest client. The details of this verification are given in App.~\ref{app:validdvnizk}.}

As next, we describe the cryptographic building blocks used in our construction. The construction itself follows in Sec.~\ref{ssec:realconstr}.

\subsection{Preliminaries}\label{sec:preliminaries}

We write $x_1,\ldots,x_n\sample X$ to denote that the values $x_1,\ldots,x_n$ are uniformly, independently sampled from a set $X$. We also write $x\sample\mathsf{X}(\ldots)$ to denote that $x$ is returned by a stochastic computation $\mathsf{X}$. We let $\ints{n}$ to denote the set $\{0,1,\ldots,2^n-1\}$.

\subsubsection{Hardness assumptions}

Let $\GG$ be a cyclic group of size $p$, with generator $g$. The \emph{discrete logarithm problem} is to find $n\in\ZZ_p$, such that $g^n=h$, for a value $h\sample\GG$. The \emph{decisional Diffie-Hellman (DDH) problem} is to distinguish tuples of the form $(g,g^x,g^y,g^{xy})$ (called \emph{Diffie-Hellman tuples}) from the tuples of the form $(g,g^x,g^y,g^z)$ for $x,y,z\sample\ZZ_p$. 
The \emph{computational Diffie-Hellman (CDH) problem} is, given $(g,g^x,g^y)$ for $x \sample \ZZ_p$ and $y \sample \ZZ_p$, come up with $g^{xy}$.

The \emph{one-more (static) computational Diffie-Hellman (CDH) problem} is, given $(g,g^x)$ for $x \sample \ZZ_p$, access to the oracle $(\cdot)^x$, and $h_0,\ldots,h_n\sample\GG$, come up with $y_0,\ldots,y_n$ satisfying $y_i = h_i^x$ while querying $(\cdot)^x$ at most $n$ times. If $n=0$, then we have the usual CDH problem.
A problem is \emph{hard} if all the efficient algorithms have at most negligible advantage (over a trivial algorithm) of solving it.

\subsubsection{ElGamal KEM and encryption scheme}
Key Encapsulation Mechanism consists of the key generation algorithm that outputs a pair of private and public key $(\sk,\pk)$; encapsulation algorithm that takes as input a public key and outputs a shared secret $ss$ and ciphertext $c$; decapsulation algorithm that on input of ciphertext $c$ and private key $\sk$ outputs a shared secret $ss$. Figure~\ref{fig:elgamal-kem} presents ElGamal KEM. 

\begin{figure}[ht!]
\begin{pchstack}
\begin{pcvstack}
\pseudocode[linenumbering, head={Key Generation:}]{
\sk \sample \ZZ_p \\
\pk=g^\sk \\
\mbox{Return $(\sk,\pk)$} 
} 
\end{pcvstack}
\pchspace[1em]
\begin{pcvstack}
\pseudocode[linenumbering, head={Encapsulation:}]{
r\sample\ZZ_p \\
\mathit{ss} = \pk^r \\
c=g^r \\
\mbox{Return $(\mathit{ss},c)$} 
}
\end{pcvstack}
\pchspace[1em]
\begin{pcvstack}
\pseudocode[linenumbering, head={Decapsulation:}]{
\mathit{ss}=c^\sk \\
\mbox{Return $\mathit{ss}$} }
\end{pcvstack}
\end{pchstack}
\caption{ElGamal KEM} \label{fig:elgamal-kem}
\end{figure}

IND-CPA security of ElGamal KEM is equivalent to the hardness of DDH in the used group $\GG$. In \emph{hashed ElGamal} KEM, the shared secret is $H'(\pk^r)$ for some hash function $H'$ that we model as a random oracle; its security is equivalent to the hardness of CDH.

A KEM can be turned into a public-key encryption scheme by combining it with a data encapsulation mechanism (DEM). In random oracle model, we may imitate the SKE2 DEM~\cite{DBLP:journals/siamcomp/CramerS03} and define the encryption of the message $m$ with the key $k$ as $\symenc{k}{m}:=(H'(k)\oplus m, H''(k,m))$ for some hash functions (random oracles) $H'$ and $H''$. The respective decryption function $\symdec{k}{(c_1,c_2)}$ computes $m\gets H'(k)\oplus c_1$, checks that $H''(k,m)=c_2$, and outputs $m$.

\subsubsection{Random oracles}
In constructions and proofs in the \emph{Random Oracle Model} (ROM)~\cite{DBLP:conf/ccs/BellareR93}, all the parties are assumed to have an access to one or several random functions $H$, where the value of the function at each point is a random variable uniformly distributed over a fixed set, and the values at different points are independent of each other. These random functions may have different codomains, e.g. the set of bit-strings of a certain length, some group $\GG$, etc., depending on the needs of the construction. The domain of a random oracle does not have to be fixed; anything encodable as a bitstring may be an input to it. In proofs, the simulator is in control of the output values of random oracles, defining them as it sees fit, as long as the distribution stays the same.

\subsubsection{(Non-interactive) zero-knowledge proofs}

A \emph{$\Sigma$-protocol} for a binary relation $R$ is a three-move protocol between two parties --- prover and verifier --- both knowing a value $x$, where the prover tries to convince the verifier that he knows some value $w$, such that $R(x,w)$ holds. The first message $\alpha$ is sent by the prover, followed by the verifier generating a fresh, independent random value $\beta$ and sending it as the second message. After the prover has sent the third message $\gamma$, the verifier runs a check on $(x,\alpha,\beta,\gamma)$ and either accepts or rejects. A $\Sigma$-protocol must have \emph{special honest-verifier zero-knowledge (ZK)}: given $(x,\beta)$, one should be able to generate $(\alpha,\gamma)$ so, that the verifier cannot distinguish them from the real protocol runs. A $\Sigma$-protocol must also be \emph{specially sound}: given $(x,\alpha,\beta_1,\beta_2,\gamma_1,\gamma_2)$ with $\beta_1\not=\beta_2$, such that the verifier accepts both $(x,\alpha,\beta_1,\gamma_1)$ and $(x,\alpha,\beta_2,\gamma_2)$, it must be possible to find $w$.

A $\Sigma$-protocol for $R$ is an (interactive) zero-knowledge proof, assuming the verifier does not deviate from the protocol. Random oracles can be used to turn $\Sigma$-protocols to non-interactive ZK (NIZK) proofs using the \emph{Fiat-Shamir (FS) transform}~\cite{fiatshamir}: instead of receiving $\beta$ from the verifier, prover himself computes it as $\beta\gets H(x,\alpha,\mathit{ctx})$, where $\mathit{ctx}$ describes the context in which the prover creates this proof. To verify the proof, the verifier recomputes $\beta$. \alisachanges{In our construction, in some proofs it is easier to assume that the part of the proof is $\beta$, not $\alpha$. The verifier then computes $\alpha$ that would satisfy the proof from $x$ and $ctx$, and checks whether $\beta = H(x,\alpha,ctx)$.}

A \emph{Designated-Verifier} NIZK (DVNIZK) proof is a NIZK proof that can convince only a single verifier. Such proofs may be cheaper to use than publicly verifiable proofs, and they may have some additional properties. Damg\r{a}rd et al.~\cite{DFN} have introduced a method (``DFN proofs'')
that applies to such $\Sigma$-protocols where $\gamma$ is computed as a linear function of $\beta$, turning them into DVNIZK proofs. In their method, the verifier generates a public-private key pair $(\ek,\vk)$ for additively homomorphic encryption. He also selects a random $\beta$ and computes $B\sample\mathcal{E}_\ek(\beta)$. Public key $\ek$ and ciphertext $B$ are given to the prover. The proof is $(\alpha,\Gamma)$, where $\alpha$ is the same as in the original $\Sigma$-protocol and $\Gamma$ is an encryption of $\gamma$, computed using $B$ and the homomorphic properties of the encryption scheme. Verifier can decrypt $\Gamma$ and perform the verification as in the original $\Sigma$-protocol. It turns out the the DFN proofs have the malleability properties that we need.

For the publicly verifiable proofs of knowledge of the exponent $r$ in an ElGamal ciphertext, we use $\Sigma$-protocols that are made non-interactive using the FS transform. As the construction of TDH1~\cite{shoupgennaro} already observed, a simple Schnorr proof of knowing $r$~\cite{schnorr} seems to be insufficient for this purpose (unless we introduce an additional extractability assumption), because the simulator will not be able to decrypt certain ciphertexts~\cite{DBLP:conf/pkc/BernhardFW16}. A more complex proof is necessary. They start from a \emph{DDH proof} $\dhp[H]{r}{g}{h}{u}{v}{\mathit{ctx}}$~\cite{chaumpedersen92} --- a NIZK proof that $\log_g u=\log_h v$ (i.e. it is not a proof \emph{of knowledge}), given in \emph{context} $\mathit{ctx}$, where $r\in\ZZ_p$ is that discrete logarithm and $H$ is a hash function, modeled as a random oracle. Denote the checking procedure by $\chp[H]{\pi}{g}{h}{u}{v}{\mathit{ctx}}$. Both proof and verification procedure are given in Figure~\ref{fig:DHP}.

\begin{figure}[ht!]
\begin{pchstack}
\pseudocode[linenumbering, head={$\dhp[H]{r}{g}{h}{u}{v}{\mathit{ctx}}$:}]{
s\sample\ZZ_p \\
\alpha \gets g^s, \quad \alpha' \gets h^s \\
\beta \gets H(g,h,u,v,\alpha,\alpha',ctx) \in \ZZ_p \\
\gamma \gets s + r \cdot \beta \\
\mbox{return $\pi \gets (\alisachanges{\beta},\gamma)$}}
\end{pchstack}
\begin{pchstack}
\pseudocode[linenumbering, head={$\chp[H]{\pi}{g}{h}{u}{v}{\mathit{ctx}}$:}]{
\alisachanges{\alpha\gets g^\gamma/  u^\beta} \\
\alisachanges{\alpha'\gets h^\gamma/ v^\beta}\\
\pcassert \alisachanges{\beta = H(g,h,u,v,\alpha,\alpha',ctx) \in \ZZ_p
}}
\end{pchstack}
    \caption{NIZK proof that $\log_g u=\log_h v$}
    \label{fig:DHP}
\end{figure}

From DDH proofs we get \emph{simulatable proofs of knowledge of exponent} $\pkexp[H,\tilde H]{r}{g}{u}{\mathit{ctx}}{\mathsf{d}}$, where $\mathsf{d}\in\{1,-1\}$ \alisachanges{is needed for exponent extraction an in described in Sec.~\ref{sec:proof}}. These prove that someone knows the value $r=\log_g u$. The construction makes use of two hash functions, both modeled as random oracles, where $H$ returns elements of $\ZZ_p$ and $\tilde H$ returns elements of $\GG$. The checking procedure $\pkchk[H,\tilde H]{\alisachanges{\pi}}{g}{u}{\mathit{ctx}}{\mathsf{d}}$ 
checks the \alisachanges{underlying} DDH proof. Both proof and verification procedure are given in Figure~\ref{fig:KNE}.

\begin{figure}[ht!]
\begin{pchstack}
\begin{pcvstack}
\pseudocode[linenumbering, head={$\pkexp[H,\tilde H]{r}{g}{u}{\mathit{ctx}}{\mathsf{d}}$:}]{
h \gets \tilde{H}(g,u,\mathit{ctx}) \in \GG \\
v \gets h^{r^\mathsf{d}} \\
\mbox{if $\mathsf{d}=1$, then  $\pi\gets\dhp[H]{r}{g}{h}{u}{v}{\mathit{ctx}}$} \\
\mbox{if $\mathsf{d}=-1$, then  $\pi\gets\dhp[H]{r}{g}{v}{u}{h}{\mathit{ctx}}$ } \\
\mbox{return $\pi' \gets (\pi,v)$}
}
\end{pcvstack}
\end{pchstack}

\begin{pchstack}
\begin{pcvstack}
\pseudocode[linenumbering, head={$\pkchk[H,\tilde H]{\pi'}{g}{u}{\mathit{ctx}}{\mathsf{d}}$:}]{
h \gets \tilde{H}(g,u,\mathit{ctx}) \in \GG \\
\alisachanges{(\pi,v) \gets \pi'}\\
\mbox{if $\mathsf{d}=1$, then } \pcassert \chp[H]{\pi}{g}{h}{u}{v}{\mathit{ctx}} \\
\mbox{if $\mathsf{d}=-1$, then } \pcassert \chp[H]{\pi}{g}{v}{u}{h}{\mathit{ctx}}
}
\end{pcvstack}
\end{pchstack}
\caption{Proof of knowledge of exponent in a group $\GG$}
\label{fig:KNE}
\end{figure}

\subsection{Our construction}\label{ssec:realconstr}

Our construction adds to the TDH1 scheme (with differently shared private key) the malleable DVNIZK proofs, and the proofs of these proofs being accepted by the server. TDH1 sets up a trapdoor with the proof of knowledge of the exponent $r$. Similarly, we need a number of trapdoors in the added proofs in order to be able to simulate a corrupted party.

\textbf{Public parameters} of $\oschiii$ contain the cyclic group $\GG$ of size $p$ with generator $g$.
They also fix a homomorphic encryption scheme $(\mathcal{K},\mathcal{E},\mathcal{D})$ for DFN proofs, where $\mathcal{K}$ generates a private and public key pair, $\mathcal{E}$ encrypts, and $\mathcal{D}$ decrypts. In this paper, that scheme will be the Paillier encryption scheme, working with moduli $N$ of bit-length $\nu\in\NN$. The public parameters moreover contain the statistical \emph{soundness} and \emph{privacy} parameters $\rho,\kappa\in\NN$ for the computations over the  integers (in the hidden-order group $\ZZ_N^*$). Typically, $\rho$ and $\kappa$ are between 80 and 256, and it is reasonable to assume that $\rho\leq \kappa\leq d$, where $d$ is the bit-length of $p$. We need $\nu$ to be several times larger than $\rho$, $\kappa$, and $d$; this will be satisfied by the Paillier encryption (where $N$ is a RSA modulus) and natural instantiations of $\GG$ (as elliptic curve groups). Let $\coins$ be the set of possible \emph{random coins} used during encryption. Finally, the public parameters contain the definitions of several random functions, modelled as random oracles, introduced below. All the sub-routines that are used in our construction are listed in the Table~\ref{tab:algorithms} to ease the understanding of the protocol.

\begin{table}[ht!]
    \centering
    \begin{tabular}{|p{2.8cm}|p{5cm}|} \hline
        $\symenc{k}{m}$ and $\symdec{k}{c}$ & encryption and decryption procedures of DEM \alisachanges{for a key $k$, a plaintext $m$ and a ciphertext $c$} \\ \hline
        $\dhp[H]{r}{g}{h}{u}{v}{\mathit{ctx}}$ and  $\chp[H]{\pi}{g}{h}{u}{v}{\mathit{ctx}}$ & proof that $\log_g u=\log_h v$, given in \emph{context} $\mathit{ctx}$, where $r\in\ZZ_p$ is that discrete logarithm\\ \hline 
        $\pkexp[H,\tilde H]{r}{g}{u}{\mathit{ctx}}{\mathsf{d}}$ and $\pkchk[H,\tilde H]{\alisachanges{\pi}}{g}{u}{\mathit{ctx}}{\mathsf{d}}$ & proof generation and verification for the proof of knowledge of exponent $r \alisachanges{= \log_g(u)}$ in context $ctx$ \\ \hline
        $\dvp{r}{\alisachanges{r'}}{\mathbf{r}}{g}{u}{\alisachanges{\alpha}}{B}{\Gamma}$ and $\dvc{\pi}{g}{u}{\alisachanges{\alpha}}{B}{\Gamma}$ & proof generation and verification for the proofs of validity of DVNIZK proof \alisachanges{$(\alpha,\Gamma)$ that uses randomness $r$, $r'$ and $\mathbf{r}$ in the context of encrypted challenge $B$} \\  \hline
        $(\mathcal{K},\mathcal{E},\mathcal{D})$ & Paillier key generation, encryption and decryption\\ \hline
    \end{tabular}
    \caption{Sub-routines used in our construction}
    \label{tab:algorithms}
\end{table}

\begin{figure}[ht!]
\centering
\begin{pcvstack}
\begin{pchstack}
\begin{pcvstack}
\pseudocode[linenumbering,head=$\kgenp{C}()$]{
\sk_1 \sample \ZZ_p; \pk_1 \gets g^{\sk_1} \\
\alisachanges{\pi_1 \gets \pkexp[H_0,{\tilde H}_0]{\sk_1}{g}{\pk_1}{\Box}{-1}}\\
com_1 \gets H_c(\pk_1\alisachanges{, \pi_1}) \\
\longrightarrow \mathrm{S}: com_1 \\
\mathrm{S} \longrightarrow: com_2 \\
\longrightarrow \mathrm{S}: \pi_1, \pk_1 \\
\mathrm{S} \longrightarrow: \pi_2, \pk_2 \\
\mathit{Verify}(com_2; \pk_2)\\
\pcassert \pkchk[H_0,{\tilde H}_0]{\pi_2}{g}{pk_2}{\Box}{-1} \\
\pk \gets \pk_2^{\sk_1}\\ \\ \\ \\ \\ \\ \\ \\
\mathrm{S} \longrightarrow: \ek,B_1,B_2 
}
\end{pcvstack}
\pchspace[1em]
\begin{pcvstack}
\pseudocode[head=$\kgenp{S}()$]{
\sk_2 \sample \ZZ_p; \pk_2 \gets g^{\sk_2} \\
\alisachanges{\pi_2 \gets \pkexp[H_0,{\tilde H}_0]{\sk_2}{g}{\pk_2}{\Box}{-1}}\\
com_2 \gets H_c(\pk_2\alisachanges{, \pi_2}) \\
\longrightarrow \mathrm{C}: com_2 \\
\mathrm{C} \longrightarrow: com_1 \\
\longrightarrow \mathrm{C}: \pi_2, \pk_2 \\
\mathrm{C} \longrightarrow: \pi_1, \pk_1 \\
\mathit{Verify}(com_1;\pk_1)\\
\pcassert \pkchk[H_0,{\tilde H}_0]{\pi_1}{g}{pk_1}{\Box}{-1} \\
\pk \gets \pk_1^{\sk_2}\\
\alisachanges{(\ek_1,\vk_1) \gets\mathcal{K}();}\\ \alisachanges{(\ek_2,\vk_2) \gets\mathcal{K}()}\\
\alisachanges{\ek = (\ek_1,\ek_2);}\\ \alisachanges{\vk = (\vk_1, \vk_2);}\\ \beta \sample \ints{\rho} \\
B_1 \sample\mathcal{E}_{\alisachanges{\ek_1}}(\beta); \\B_2\sample\mathcal{E}_{\alisachanges{\ek_2}}(\beta) \\
\longrightarrow \mathrm{C}: \ek, B_1, B_2\\
}
\end{pcvstack}
\end{pchstack}
\begin{pchstack}
\pchspace[1em]
\begin{pcvstack}
\pseudocode{
\sendmessage{<-}{length=7cm, top=Prove that $\mathcal{D}_{\alisachanges{\vk_1}}(B_1)\in\ints{\rho}$}\\
\sendmessage{<-}{length=7cm, top=Prove that $\mathcal{D}_{\alisachanges{\vk_1}}(B_1) \equals \mathcal{D}_{\alisachanges{\vk_2}}(B_2)$}
}
\end{pcvstack}
\end{pchstack}
\begin{pchstack}
\begin{pcvstack}
\pseudocode{
pk_1 \gets (\pk,\ek,B_1,B_2)\\
\pcreturn \sk_1, pk_1
}
\end{pcvstack}
\pchspace[5.0em]
\begin{pcvstack}
\pseudocode{
pk_2 \gets (\pk,\ek,B_1,B_2)\\
\pcreturn (\sk_2,\vk,\beta), pk_2
}
\end{pcvstack}
\end{pchstack}
\end{pcvstack}
\caption{Key generation for client and server in our construction}\label{fig:keygen}
\end{figure}

\begin{figure*}[ht!]
\centering
\begin{pchstack}
\begin{pcvstack}
\pseudocode[linenumbering,head=$\enca(m\comma{} \pk\comma{} \alisachanges{\ek_1 \comma{} \ek_2}\comma{} B_1\comma{} B_2)$]{
r\sample\ZZ_p\\
u\gets g^{r}\\
c_2\gets \symenc{\pk^r}{m}\\
r_1\sample\ints{\rho+d+\kappa}\\
\mathbf{r}_\iota\sample \coins\\
\alpha_1\gets g^{r_1\bmod p}\\
\Gamma_\iota\gets \mathcal{E}_{\alisachanges{\ek_{\iota}}}(r_1;\mathbf{r}_\iota)\cdot B_\iota^{r}\\
\pi\gets\pkexp[H_1,{\tilde H}_1]{r}{g}{u}{\alpha_1,\Gamma_1,\Gamma_2}{1}\\
\pi_\iota\gets\dvp{r}{r_1}{\mathbf{r}_\iota}{g}{u}{\alpha_1}{B_\iota}{\Gamma_\iota}\\
c_1\gets (u,\alpha_1,\Gamma_1,\Gamma_2,\pi,\pi_1,\pi_2)\\
\pcreturn c_1, c_2\pcskipln\\[5mm]
\iota\in\{1,2\}
}
\end{pcvstack}
\pchspace
\begin{pcvstack}
\pseudocode[linenumbering,head=$\decp{C}(c_1\comma{} c_2\comma{} \pk_1\comma{} \sk_1\comma{} \pk\comma{} \alisachanges{\ek_1 \comma{} \ek_2}\comma{} B_1\comma{} B_2)$]{
(u,\alpha_1,\Gamma_1,\Gamma_2,\pi,\pi_1,\pi_2) \gets c_1\\
\pcassert \pkchk[H_1,{\tilde H}_1]{\pi}{g}{u}{\alpha_1,\Gamma_1,\Gamma_2}{1}\\
\pcassert \dvc{\pi_\iota}{g}{u}{\alpha_1}{B_\iota}{\Gamma_\iota}\\
z\sample\ZZ_p, \ z'\sample \ints{\rho+2d+2\kappa}\\
\mathbf{r}'_\iota\sample \coins\\
u'\gets u^z\\
\alpha'_1\gets \alpha_1^z\cdot g^{z'\bmod p}\\
\Gamma'_\iota\gets \Gamma_\iota^z\cdot\mathcal{E}_{\alisachanges{\ek_{\iota}}}(z'; \mathbf{r}'_\iota)\\
\pi'\sample\pkexp[H_2,{\tilde H}_2]{\sk_1}{g}{\pk_1}{u',\alpha'_1,\Gamma'_1,\Gamma'_2}{-1}\\
\longrightarrow \mathrm{S}: u',\alpha'_1,\Gamma'_1,\Gamma'_2,\pi'
\\
\mathrm{S}\longrightarrow: w,\pi''\\
\pcassert \chp[H_3]{\pi''}{\pk_1}{u'}{\pk}{w}{\Box}\\
\pcreturn \symdec{w^{\sk_1/z}}{c_2}
}
\end{pcvstack}
\pchspace
\begin{pcvstack}
\pseudocode[linenumbering,head=$\decp{S}(\sk_2\comma{} \pk_1\comma{} \pk\comma{} \alisachanges{\vk_1 \comma{} \vk_2}\comma{} \beta)$]{
\mathrm{C}\longrightarrow: u',\alpha'_1,\Gamma'_1,\Gamma'_2,\pi'
\\
\gamma'\gets \mathcal{D}_{\alisachanges{\vk_1}}(\Gamma'_1)\\
\pcassert \gamma'=\mathcal{D}_{\alisachanges{\vk_2}}(\Gamma'_2)\\
\pcassert g^{\gamma'} = \alpha'_1 \cdot (u')^\beta\\
\pcassert \pkchk[H_2,{\tilde H}_2]{\pi'}{g}{\pk_1}{u',\alpha'_1,\Gamma'_1,\Gamma'_2}{-1}\\
w\gets (u')^{\sk_2}\\
\pi''\gets\dhp[H_3]{\sk_2}{\pk_1}{u'}{\pk}{w}{\Box}\\
\longrightarrow \mathrm{C}: w,\pi''\\
\pcreturn \top
}
\end{pcvstack}
\end{pchstack}
\caption{Encryption and decryption in our construction}\label{fig:enc-dec}
\end{figure*}

\textbf{Key generation} of $\oschiii$ is given in Fig.~\ref{fig:keygen}. During the key generation, the client and the server respectively select private key shares $\sk_1\sample\ZZ_p$ and $\sk_2\sample\ZZ_p$, compute public key shares $\pk_i=g^{\sk_i}$ (line 1), and exchange the latter values using hash commitments (lines 2--4). Beside $\pk_i$, the client and server also exchange the proofs $\pkexp[H_0,{\tilde H}_0]{\sk_i}{g}{\pk_i}{\Box}{-1}$ that they know their respective private keys (lines 5--7), \lizachanges{this  prevents  a malicious  party  from  choosing their public key share based on the share of the honest party without knowing corresponding exponent.} They will verify the proofs (lines 8--9), and define combined public key $\pk=\pk_{3-i}^{\sk_i}$ (line 10), i.e. the private key is \emph{multiplicatively}, not additively shared.

Having set up $\pk$, the parties continue with setting up the public and private values for DVNIZK proofs. \lizachanges{DVNIZK proofs are needed for the encrypting party to prove that they know the randomness used to generate ElGamal ciphertext. Moreover, chosen DVNIZK proofs allow client to blind the proof before sending it in the decryption query to the server (who is acting as a designated verifier).} The server generates \alisachanges{two key pairs $(\ek_1,\vk_1)$ and $(\ek_2,\vk_2)$} for the homomorphic encryption scheme, where \alisachanges{$\ek_i$} is the public key and \alisachanges{$\vk_i$} the private key (line 11). The set-up also consists of public ciphertexts $B_1\sample\mathcal{E}_{\alisachanges{\ek_1}}(\beta)$ and $B_2\sample\mathcal{E}_{\alisachanges{\ek_2}}(\beta)$ (line 12), where $\beta\sample\ints{\rho}$ is generated and kept secret by the server. \lizachanges{The necessity of generating two instances comes from the security and we provide detailed explanation in Section~\ref{sec:proof}.} Server sends $\ek\alisachanges{=(\ek_1,\ek_2)}$ and $B_1$ and $B_2$ to the client. Finally, the set-up for DFN proofs consists of the server proving that $B_1$ indeed encrypts a value at most $\rho$ bits long and proving that $B_1$ and $B_2$ encrypt the same value. 

The public key is $(\pk,\alisachanges{(\ek_1,\ek_2)},B_1,B_2)$, allowing an encryptor to create the ElGamal ciphertext with the public key $\pk$, and the two DVNIZK proofs of knowledge of the random exponent of the ciphertext, using $(\alisachanges{\ek_1},B_1)$ and $(\alisachanges{\ek_2},B_2)$ (the NIZK proof of knowledge of the exponent, as well as the proofs of the DVNIZK proofs being correct, are standard FS transforms of $\Sigma$-protocols). Client keeps $\sk_1$ as his keyshare. Server keeps $(\sk_2,\alisachanges{(\vk_1,\vk_2)},\beta)$ as his keyshare. The both also keep $\pk_1,\pk_2$.

\textbf{Encryption} and \textbf{decryption} are given in Fig.~\ref{fig:enc-dec}. As the operations with $B_1$ and $B_2$ are mostly similar, we abbreviate the write-up by introducing the convention that the parameter $\iota$ ranges over $\{1,2\}$ and the operations of defining values $X_1$ and $X_2$ are given only once by stating how $X_\iota$ is computed. 

The encryption of the message $m$ starts by computing its ``standard'' ElGamal ciphertext $(u,c_2)$ (lines 1--3). In line 4--7, we compute the two DVNIZK proofs of the encryptor knowing $r$. \lizachanges{Those proofs can only be verified by the server during the decryption process.} The value $\alpha_1$ is the first message of both proofs; it is computed as in a Schnorr proof of knowledge, raising the generator $g$ to a random power $r_1$. The second message is the server's secret $\beta$, and the third message $\Gamma_\iota$ is an encryption of $\gamma=r_1+\beta r$. 

In line 8, we compute the (usual) NIZK proof that the encryptor knows $r$. The same proof, with the same trapdoor is present in TDH1. \lizachanges{This proof can be verified by the client.} Finally, in line 9 of $\enca$ we compute the proofs $\pi_\iota$ ($\iota\in\{1,2\}$) that an honest server will accept the designated verifier proofs $(\alpha_1,\Gamma_\iota)$, while verifying them in the manner described in Sec.~\ref{sec:preliminaries}. This proof is given using standard cryptographic techniques; we present it in App.~\ref{app:validdvnizk}. No trapdoor is embedded in $\pi_\iota$. \lizachanges{The proof is required for the client not to engage in decryption of an incorrect ciphertext. If the proof does not verify on the client side, the client does not initiate protocol with the server.} The final ciphertext is $[c_1=(u,\alpha_1,\Gamma_1,\Gamma_2,\pi,\pi_1,\pi_2),c_2]$.


One has to be careful with the moduli: while the exponents of $g$ are naturally taken modulo $p$, the computations under encryption are done modulo the RSA/Paillier modulus $N$. As $N$ is much larger than $p$, we can just make sure that certain computations do not overflow. In particular, we want $\gamma=r_1+\beta r$ be less than $N$. We also want $r_1$ to hide $\beta r$. Obtaining the perfect security may be hard, but we can get statistical security. The value $\beta r$ is $\rho+d$ bits long. Hence we let $r_1$ to be $\rho+d+\kappa$ bits long (which must still be less than the size $\nu$ of $N$). This is similar to homomorphic commitments to integers~\cite{Damgard-Fujisaki-AC02,eprint-2006-044}.


During the \textbf{decryption}, the client first verifies two of three proofs in lines 2--3. He will then construct the blinded encapsulation $u'$ (line 6) by raising $u$ to a random exponent $z$. Hence $u'$ is independent of $u$. In lines 7--8, the client will then blind the proof $(\alpha_1,\Gamma_\iota)$ of knowledge of $r=\log_g u$ to get the proof $(\alpha'_1,\Gamma'_\iota)$ of knowledge of $rz=\log_g u'$. $(\alpha_1^z,\Gamma_\iota^z)$ is such a proof, as long as the plaintext $z\gamma=\mathcal{D}_{\alisachanges{\vk_\iota}}(\Gamma_\iota^z)$ is not reduced modulo $N$. That reduction indeed does not happen, because $z\gamma$ is at most $\rho+2d+\kappa\leq\nu$ bits long.

However, $z\gamma$ may leak something about $z$ to the server, once the latter has decrypted $\Gamma_\iota^z$ and obtained it. Indeed, $z$ is a factor of $z\gamma$. Hence we will further additively mask $(\alpha_1^z,\Gamma_\iota^z)$, using $z'$. We get statistical security if we let $z'$ to be at last $\kappa$ bits longer than $z\gamma$.

The client proceeds with generating proof $\pi'$ that they know their private key in the context of blinded ciphertext that is being decrypted (line 9). \lizachanges{This proof authenticates client to the server, demonstrating that they know the private key share.} The clients sends $u',\pi'$ and the malleated DVNIZK proofs to the server (line 10), who verifies received proofs (line 2--5 in $\decp{S}$). The rest of the protocol is straightforward, computing $\pk^r = (((u')^{\sk_2})^{\sk_1})^{1/z}$ from $u'$. The server does the first exponentiation in line 6 of $\decp{S}$ and sends it back to client (line 8), accompanying it with a proof (line 7) that he performed the exponentiation correctly. After verifying this proof (line 12 of $\decp{C}$) , the client performs the second and third exponentiation, and uses the obtained $\pk^r$ to find the plaintext $m$ from $c_2$.

\peeterchanges{In the presented protocol, client contacts the server, which will then use its private key share. We do not include (neither here nor in the definition of $\mathcal{F}$ in Fig.~\ref{fideal}) the details on how the server is \emph{activated}. Indeed, if the server is always responsive, then an adversary with client's keyshare will be able to decrypt ciphertexts by masquerading as the client to the server. Formally, the activation of the server is decided in the environment. In practice~\cite{splitkey,kirss-et-al}, client's keyshare is stored encrypted with a PIN, with guesses of the PIN impossible to verify without contacting the server. A mechanism for detecting the existence of several clones of keyshare is also used~\cite{Sarr19}.}

\section{Security proofs}\label{sec:proof}

We begin by introducing an augmented version of the one-more CDH problem, whose hardness is a stronger assumption than hardness of the standard one-more CDH problem. 
In the \emph{one-more (static) CDH problem with tests}, one is given $(g,g^x)$ with $x\sample\ZZ_p$, random $h_0,\ldots,h_n\in\GG$, the oracle $(\cdot)^x$, and the oracle $\chck(\cdot)$,  where $\mathsf{Check}(h)$ returns $i$ such that $h=h_i^x$, and $\bot$ if there is no such $i$. The goal of the solver is to invoke $\chck(h_i^x)$ for all $i\in\{0,\ldots,n\}$, while making at most $n$ calls to the oracle $(\cdot)^x$. The calls to $\chck$-oracle are not restricted in number. While the $\chck$-oracle may give the solver some extra power in the Standard Model, Bauer et al.~\cite[Thm.~5]{testingoracle} show that nothing is gained in the Algebraic Group Model (AGM)~\cite{algebraicgroup}.

\lizachanges{Our scheme is instantiated with the simulatable proofs of knowledge of exponent \alisachanges{$r = \log_g(u)$ (Fig.~\ref{fig:KNE})} that additionally allow the simulator to raise an \alisachanges{arbitrary element $z \in \GG$} of its choice to the power of $r^\mathsf{d}$ where $\mathsf{d}\in\{1,-1\}$. The simulator has to choose the \alisachanges{value $z$} at the time the adversary computes the corresponding proof. If the simulator wants to obtain $z^{r^\mathsf{d}}$, it will follow the instructions from Figure~\ref{fig:exponent}. \alisachanges{The quantity $z^{r^{\mathsf{d}}}$ can be computed as soon as the simulator gets the proof.}}
\begin{figure}[ht!]
\begin{tcolorbox}[colback=white,arc=0.3mm, boxrule=0.3mm,fontupper=\normalsize]
\underline{Raising $z \in \GG$ to the power $r^{\mathsf{d}}$ for $\mathsf{d}\in\{1,-1\}$:}
\begin{enumerate}
    \item generate $t\sample \ZZ_p$
    \item program $\tilde H_i$ to return $h = z^{1/t}$ when the adversary queries it with $g,u,\mathit{ctx}$
    \item \alisachanges{at any point later, after obtaining the proof $(\pi,v)$ from the adversary, compute} $z^{r^\mathsf{d}}=v^t$. \alisachanges{If the proof $\pi$ is valid, then $v = h^{r^{\mathsf{d}}}$.} 
\end{enumerate}
\end{tcolorbox}
    \caption{Raising $z \in \GG$ to the power of $r^{\mathsf{d}}$ for $\mathsf{d}\in\{1,-1\}$}
    \label{fig:exponent}
\end{figure}

\lizachanges{Additionally, in proofs of security against the malicious server, the simulator may need to know the value $\beta$. Hence we give the proof of $\beta\in\ints{\rho}$ in a manner that allows the simulator to extract it. The proof can be given using standard cryptographic techniques, we describe it in App.~\ref{app:rangeproof}. We believe that as key generation is a seldomly-performed operation, its high efficiency is relatively less important, as long as it can be performed in reasonable time. Moreover, in proofs of security against malicious client, the simulator may need to know the witnesses that the client has used in the creation of DFN proofs. While Damg\r{a}rd et al.~\cite{DFN} introduced a \emph{knowledge assumption} stating that such extraction is possible, we choose to add more components to the ciphertexts, allowing us to assume the hardness of DDH, one-more CDH and DCRA only. Thus, during the key generation the server also computes $B_2\sample\mathcal{E}_{\alisachanges{\ek_2}}(\beta)$. When the simulator is acting as the server, it will be able to make $B_1$ and $B_2$ contain different values \alisachanges{$\beta_1 \neq \beta_2$}, faking the plaintext equality proof. It will then be able to use the special soundness of the $\Sigma$-protocols to find the witness. Again, the proof can be given using standard cryptographic techniques, see App.~\ref{app:paillierpaillier}.}

\begin{theorem}\label{thm:main}
The protocol set $\oschiii$ in Figures~\ref{fig:keygen} and~\ref{fig:enc-dec} securely implements functionality $\fideal$ in presence of malicious static adversary under one-more CDH assumption with tests.
\end{theorem}
In this section, we give a proof sketch of Theorem~\ref{thm:main}. The full proof can be found in App.~\ref{app:fullproof}.

In order to show universal composability security of our threshold decryption protocol, we construct a simulator $\simm$ such that for each adversary $\adv$ attacking the protocol, the environment $\zenv$ cannot distinguish whether it is interacting with the real protocol and the adversary $\adv$, or the ideal functionality $\fideal$ and the ``ideal'' adversary (consisting of $\adv$ and $\simm$). \lizachanges{We start with defining $\simm$ and how it interacts with the ideal functionality and the adversary in case of different parties being corrupted.}  

\alisachanges{
\textbf{Initialization:}}

\alisachanges{We assume that the corruptions are known already in the initialization phase. There is a parameter $n$ denoting the number of encryptions considered in this session. From an external challenger, $\simm$ receives as an input $\pk = g^{\sk}$ (where the $\sk$ is only known to the challenger) and also the following:}
\begin{itemize}
\item If the client is \emph{corrupted}, $\simm$ receives $u_0,\ldots,u_n$ for some parameter $n$, such that $(g,\pk,u_0,\ldots,u_n)$ is an instance of a one-more CDH problem. $\simm$ also gets access to an oracle $(\cdot)^{\sk}$. \alisachanges{Intuitively, if the number of decrypted ciphertexts exceeds the number of backdoor decryption queries allowed by $\fideal$, then the $\simm$ finds solution to the one-more CDH problem.}
\item If the client is \emph{honest}, $\simm$ receives $\pk$ and $u_0$, so that $(g,\pk,u_0)$ is an instance of a CDH problem, i.e. one-more CDH problem with $n=0$. $\simm$ takes $u_1=\cdots=u_n \gets u_0$. \alisachanges{Intuitively, breaking the IND-CCA property of the scheme would mean breaking the CDH problem.}
\end{itemize}
In addition, in both cases, $\simm$ gets access to the oracle $\chck(\cdot)$.

\lizachanges{We proceed with arguing that the encryption and decryption performed by the ideal functionality interacting with the $\simm$ is indistinguishable from the real protocol. If the \alisachanges{environment} is able to distinguish between the real protocol and the ideal protocol, \alisachanges{then the simulator has enough information to solve one-more CDH problem that it received as an input.}}

\alisachanges{The simulator prepares tables for storing the random oracle data, and in the background it runs processes that are constantly observing accesses of the adversary $\adv$ and the environment $\zenv$ to these oracles, programming them on demand to extract certain values as described further.}

\alisachanges{\textbf{Key  generation:}
\begin{itemize}
\item Use random oracle $H_c$ to extract the corrupted party's share of $\pk$ and the proof $\pi$ from the commitment.
\item Use knowledge extraction w.r.t. $\pi$ to adjust the honest party's share of $\pk$ to the input $\pk$ and the share $\sk_i$ of the corrupted party (by applying the procedure of Fig.~\ref{fig:exponent} to the random oracles $H_0$ and $\tilde{H}_0$ ).
\item Simulate the proofs related to $\beta$. If the client is corrupted, $\simm$ comes up with $\beta_1$ and $\beta_2$ such that $\beta_1 \neq \beta_2$ to be able to apply special soundness later. As the simulator has access to trapdoors, the proofs of equality can be simulated even if $\beta_1 \neq \beta_2$.
\end{itemize}}

\alisachanges{\textbf{Encryption:}
\begin{itemize}
\item Whenever $\fideal$ is used to encrypt a message, it asks for a ciphertext from $\simm$, who constructs the ciphertext from the next challenge. In particular, it computes $u = (u_j)^r$ for a challenge $u_j$ instead of taking $u = g^r$ for a known $r$. Also, it outputs a random $c_2$ instead of encrypting the true message (which it does not know) so that the random oracles of $H'$ and $H''$ could be programmed later. All related proofs can be simulated.
\item For any ciphertext encrypted outside of $\fideal$, a valid proof needs to be constructed for a later successful decryption, for which oracles $H_1$ and $\tilde{H}_1$ needs to be accessed. This allows $\simm$ to prepare random oracles $H_1$ and $\tilde{H_1}$ to compute $\pk^r$ as in Fig.~\ref{fig:exponent}.
\end{itemize}}

\begin{figure}[ht!]
\begin{pcvstack}
\pseudocode[linenumbering,head=$\ver^{H_1,\tilde{H_1},B_1,B_2}(c_1\comma{}c_2\comma{}c_3)$]{
(u, \alpha_1,\Gamma_1,\Gamma_2,\pi,\pi_1,\pi_2) \gets c_1 \\
\pcassert \pkchk[H_1,{\tilde H}_1]{\pi}{g}{u}{\alpha_1,\Gamma_1,\Gamma_2}{1}\\
\pcassert \dvc{\pi_\iota}{g}{u}{\alpha_1}{B_\iota}{\Gamma_\iota}\\
\pcreturn \mathsf{true}
}
\end{pcvstack}
\caption{Ciphertext correctness verification function $\ver$ (delegated by the simulator to $\fideal$). Here $H_1$ and $\tilde{H_1}$ are provided as tables.
}
\label{fig:verify-alg}
\end{figure}

\begin{figure}[ht!]
\begin{pcvstack}
\pseudocode[linenumbering,head=$\dec^{\knetable{1},H',H''}(c_1\comma{}\alisachanges{c_2})$]{
(u, \alisachanges{\ldots,\pi},\ldots) \gets c_1 \\
\pcif (m,((u,\ldots),\alisachanges{c_2})) \pcin\ \mathsf{T} \pcthen\\
\t\pcreturn m\\
\alisachanges{(\ldots,v) \gets \pi}\\
\pcget (u,\alisachanges{v},t)\ \pcfrom \knetable{1}\\
k \gets v^{t}\\
\alisachanges{(c'_2,c''_2) \gets c_2}\\
\pcassert c''_2 = H''(k, c'_2)\\
\text{compute }m \gets H'(k) \oplus c'_2\\
\pcreturn m
}
\end{pcvstack}
\caption{Decryption function $\dec$ (delegated by the simulator to $\fideal$). If only the proof part of $c$ is different from some existing ciphertext $c'$, take $(m,c')$ from $\mathsf{T}$. Otherwise, decrypt as in the real protocol. Although $\fideal$ does not know the secret key, \alisachanges{it can compute $k = v^t$ using the table $\knetable{1}$ that contains all the entries $(u,v,t)$ for which $\simm$ has made precomputations of Fig.~\ref{fig:exponent}. It extracts the entry that corresponds to $u$ and $v$} (if there is no such entry, return $\bot$).  Here $H'$ and $H''$ are provided as tables.
}
\label{fig:decryption-alg}
\end{figure}

\alisachanges{\textbf{Decryption (honest client):}
\begin{itemize}
\item $\fideal$ waits for a verification function $\ver$ from $\simm$, who chooses it as the initial knowledge extraction and the DVNIZK verification on behalf of the client. As it needs to evaluate $H_1$ and $\tilde{H}_1$, the contents of these hash tables are delivered as a part of procedure $\ver$ (Fig.~\ref{fig:verify-alg}). The quantities $B_1$ and $B_2$ are also delivered as a part of $\ver$.
\item As $\simm$ does not know the true ciphertext, interaction with the server is simulated for a ciphertext generated from a \emph{random plaintext}. Due to the statistical blinding, the environment detects the difference with the negligible probability.
\item $\fideal$ waits for a decryption function $\dec$ from $\simm$, who chooses it in such a way that $\fideal$ can combine the proof $\pi$ with the data of the $H_1$ and $\tilde{H}_1$ to get $\pk^r$. For this, $\simm$ provides a table $\knetable{1}$ of all triples $(u,v,t)$ obtained so far for $H_1$ and $\tilde{H}_1$ as in Fig.~\ref{fig:exponent}, so that $\fideal$ can take the one that corresponds to $u$ and $v$. The function $\dec$ (Fig.~\ref{fig:decryption-alg}) takes into account that different ciphertexts \emph{can} depend on each other as far as only the proof part is different, so it lets $\fideal$ take a record $(m,c)$ from its internal table even if only $u$ and $c_2$ match, but the proofs look different (as far as they are valid).
\item For any ciphertext generated \emph{inside} $\fideal$, the attacker could notice inconsistency of decryption only if it manages to construct a ciphertext $((u^*,\ldots),c^*_2)$ which would decrypt to something that is related to a message that is encrypted with $\pk^r = u^{\sk}$ where $u$ is a CDH challenge, but $u^{*} \neq u$ or $c_2^* \neq c_2$. To construct such a ciphertext the attacker would need to solve CDH.
\end{itemize}}

\alisachanges{
\textbf{Decryption (corrupted client):}
\begin{itemize}
\item The simulator gets from $\adv$ a blinded ciphertext and needs to come up with a response $w$ and a proof $\pi'$. The response should be valid for a valid ciphertext.
\begin{itemize}
\item For messages generated outside of $\fideal$, using special soundness, $\simm$ can compute $r' = \log_g(u')$ from a valid sigma-proof, which allows to compute $w = \pk_2^{r'}$.
\item For messages generated inside $\fideal$, it is impossible to extract the exponent, so $\simm$ needs to compute $w = (u')^{\sk_2} = (u'^{1/\sk_1})^{\sk}$. For this, it needs to access the oracle $(\cdot)^\sk$. To obtain $u'^{1/\sk_1}$ from $u'$, $\simm$ uses Fig.~\ref{fig:exponent} to prepare random oracles $H_2$ and $\tilde{H_2}$ to compute this value at the point when a valid proof is created (the value $u'$ is a part of the context which is an input to $\tilde{H}_2$).
\item In general, homomorphic encryption allows to obtain a valid ciphertext as a linear combination of ideal ciphertexts. However, since $\simm$ has used the same $\gamma$ in $\Gamma_1$ and $\Gamma_2$ of the challenge ciphertexts, special soundness allows to extract the free term $r'$, splitting $u' = u'_1 \cdot u'_2$ where $u' = g^{r'}$ and $u'_2 = \prod_j u_j^{z_j}$ is a linear combination of challenges. The simulator can compute $w = \pk_2^{r'} \cdot (u'_2)^{\sk}$, thus ensuring that $(\cdot)^{\sk}$ is only applied to the linear combinations of challenges and is not misused e.g. to break some proof.
\end{itemize}
\item In any case, there will be no more oracle $(\cdot)^{\sk}$ called than there are decryptions approved by the server. At any point of time when the adversary attempts to actually decrypt a ciphertext (it may happen later, after the decryption has ended), the simulator has control over it as it controls $H'$ and $H''$ that are valuated for a valid decryption. The simulator programs $H'$ and $H''$ so that it would decrypt to the true message. If it is one of the previous ciphertexts by $\fideal$ (this fact can be verified using $\chck(\cdot)$ oracle), the simulator can use the "backdoor" decryption of $\fideal$ to get the plaintext. If the adversary manages to obtain more plaintexts than there have been decryption sessions (and hence allowed "backdoor" decryptions), $\simm$ has solved the one-more-CDH problem.
\end{itemize}}

\peeterchanges{\section{Efficiency}}
The performance of the encryption scheme can most probably be improved, perhaps by relying on different hardness assumptions. But even currently, the performance is very satisfactory for the intended application --- decrypting credentials before their use. We have implemented $\oschiii$ encryption and decryption in Python on top of the PyCryptodome cryptographic library, as well as the \emph{python-paillier} library~\cite{PythonPaillier} for the Paillier encryption. We use the elliptic curve group P-256 as $\GG$, and 3072-bit RSA modulus for Paillier. The running times on a laptop with an Intel\textregistered{} Core\texttrademark{} i5-10210U CPU and 16GB RAM are 1140 ms for $\enca$, 563 ms for $\decp{C}$, and 153 ms for $\decp{S}$. The sizes of the messages sent from the client to the server and back are 11.5KB and 5.5KB, respectively.

\peeterchanges{We can count the size our ciphertexts as follows. Ignoring the symmetrically encrypted payload $c_2$ (Fig.~\ref{fig:enc-dec}), the component $c_1$ consists of two elements of $\GG$ (each represented by $d+1$ bits), two Paillier ciphertexts, and three proofs $\pi,\pi_1,\pi_2$, where the first is different from the other two. The proof $\pi$ consists of a group element and two numbers of length $\rho$ and $d$, respectively. The proof $\pi_\iota$ (Fig.~\ref{fig:validdvnizk}) consists of a $\rho$-bit hash value, two numbers of length $(\rho+d+\kappa)$ and $(2\rho+d+2\kappa)$, respectively, and one value that can serve as the random coins for the homomorphic encryption. We have chosen the security parameter determining the privacy of the protocol as $\kappa=128$. It is also reasonable to take the integrity parameter as $\rho=80$. Moreover, the size of group elements is $d=256$. Considering that Paillier ciphertexts are 6144 bits long and Paillier coins (which are elements of $\ZZ_N$ for the used RSA modulus $N$) are 3072 bits long, the total size of the ciphertext $c_1$ is 21714 bits.}

\peeterchanges{Most of the size of $c_1$ is taken up by two Paillier ciphertexts and two Paillier coins. There exist additively homomorphic encryption schemes with smaller ciphertext sizes. The Castagnos-Laguillaumie (CL) encryption system~\cite{cl-encryption} has ciphertexts of size 4166 bits at the same security level~\cite{castagnos-habilitation}, with coins of ca. 1955 bits. The homomorphism is according to a smaller modulus, but sufficient for our purposes, where we assume that there is no modular reduction during the operations inside the ciphertexts. Using CL encryption, our ciphertexts would be ca. 15500 bits in length. Similar sizes may be achievable with Joye-Libert cryptosystem~\cite{jl-encryption}.}

\peeterchanges{We can compare these sizes with the ciphertext sizes in the schemes of Green~\cite{green} and Blazy et al.~\cite{blazy}, both of which are pairing-based constructions. In the latter, the ciphertexts consist of 14 elements of the first source group, 15 elements of the second source group, and 4 elements of the target group. Assuming the use of the BLS12-381 curve~\cite{bls12381}, we estimate that the elements of these groups take ca. $384$, $2\cdot 384$, and $12\cdot 384$ bits to represent, respectively. In this case, the size of the ciphertext is ca. 35328 bits. Green's scheme is given in the setting of \emph{symmetric pairings}; the ciphertext consists of 25 elements of the source group and two numbers (used as exponents). We believe that in asymmetric setting, most components of the ciphertext can be elements of the first source group. With BLS12-381 curve, the size of the ciphertext should be in the range of 10--11 kilobits. As we discussed above, the goals of Green's scheme are different from ours.}

\section{Conclusions}\label{sec:conclusions}

We have discussed the existing security definitions for the blind assisted and/or threshold decryption, demonstrated their shortcomings, proposed new ones, and instantiated them with the $\oschiii$ scheme. The security of the scheme relies on well-known hardness assumptions, it combines existing building blocks in somewhat novel manner, and its security is proved using very recent techniques.

Our security definition is given in the universal composability model. We have used this model to be able to capture the details of of the security definition, capturing both the IND-CCA security and privacy notions. \peeterchanges{We have also given the first ever treatment of threshold decryption in the UC model, arguing what must be the interface that the ideal functionality offers to the simulator / the ideal adversary. We show how this interface changes when the decryption is privacy-preserving.}

\textbf{Acknowledgement.} This research has been supported by Estonian Research Council, grant No. PRG1780.

\bibliographystyle{IEEEtran}
\bibliography{refs}

\appendices

\section{Subprotocols for key generation}

\subsection{Proving equality of Paillier encryption and Pedersen commitment}\label{app:paillierpedersen}

This proof is a subprotocol in protocols given below. It can be constructed from standard building blocks, as shown by Lindell et al.~\cite[Sec.~6.2.4]{cryptoeprint:2018/987}. In our setting, the server has presented a Paillier ciphertext $E=\mathcal{E}_\ek(x)$ and a Pedersen commitment $C=g^x h^z\in\GG$ to the client, where the server knows Paillier private key $\vk$, the value $x$, the exponent $z$, and also the random coins used to create $E$. The value $g\in\GG$ is the generator of $\GG$, while $h$ is another element of $\GG$, such that neither the client nor the server know $\log_g h$.

We use this subprotocol unchanged from~\cite{cryptoeprint:2018/987}, and will not copy it here.

\subsection{Simulatable proof of equality of plaintexts in Paillier encryptions}\label{app:paillierpaillier}
\peeterchanges{At some point, server has sent two Paillier ciphertexts $E_1$ and $E_2$ to the client, and wants to prove him that $x_1=x_2$, where $x_i=\mathcal{D}_{\vk_i}(E_i)$. In order to do this, the server creates a Pedersen commitments $C_1,C_2$ to $x_1$ and $x_2$ --- it computes $C_i\gets g^{x_i}h^{z_i}$, where $z_1,z_2\sample\ZZ_p$. The server proves that the ciphertext $E_i$ contains the same value as $C_i$ (App.~\ref{app:paillierpedersen}) for $i\in\{1,2\}$. The server also makes, and client verifies the proof $\pi\leftarrow\pkexp{z}{h}{C}{\Box}{1}$, where $z=z_1-z_2$ and $C=C_1/C_2=g^{x_1-x_2}h^z$. The latter proof implies that server knows $\log_h C$, meaning that $C$ must have been computed as $g^0h^z$, because server does not know $\log_g h$.}

This proof is simulatable, because the simulator knows $\log_g h$. It can thus produce $\pi$ (or run the interactive version of it) even if $C$ is not a commitment to $0$.

\subsection{Disjunction of $\Sigma$-protocols}\label{app:sigmadisjunct}

Suppose there are $\Sigma$-protocols $\Sigma_1$, $\Sigma_2$ for two relations $R_1$ and $R_2$, and Prover and Verifier have an instance $x_1$, $x_2$ for each of them. Let Prover also have witness $w_\iota$ for one of them, i.e. $\iota\in\{1,2\}$. Prover wants to show that it has this $w_\iota$, but wants to disclose neither it nor the index $\iota$ to Verifier. Let the challenge messages in $\Sigma_1$ and $\Sigma_2$ belong to the same set, and let this set also have the structure of an Abelian group. The protocol for such disjunction is the following.
\begin{enumerate}
    \item Prover generates the ``challenge'' $\beta_{3-\iota}$.
    \item Using the zero-knowledge property of $\Sigma_{3-\iota}$, Prover generates $(\alpha_{3-\iota},\gamma_{3-\iota})$, such that $(x_{3-\iota},\alpha_{3-\iota},\beta_{3-\iota},\gamma_{3-\iota})$ would pass Verifier's final check in $\Sigma_{3-\iota}$.
    \item Prover generates the message $\alpha_\iota$ in the protocol $\Sigma_\iota$, using $x_\iota,w_\iota$ if necessary.
    \item Prover sends $(\alpha_1,\alpha_2)$ to Verifier.
    \item Verifier generates the challenge $\beta$ and sends it to Prover.
    \item Prover defines $\beta_\iota\gets\beta-\beta_{3-\iota}$.
    \item Prover constructs the message $\gamma_\iota$ in protocol $\Sigma_\iota$, using $\beta_\iota$ as the challenge.
    \item Prover sends $(\beta_1,\beta_2,\gamma_1,\gamma_2)$ to Verifier.
    \item Verifier performs the checks in both $\Sigma_1$ and $\Sigma_2$, using $\beta_1,\beta_2$ as the challenges. Also checks that $\beta_1+\beta_2=\beta$.
\end{enumerate}
Such protocol is again a $\Sigma$-protocol. Both the zero-knowledge simulation, and extraction algorithms for the composed protocol can easily be constructed from the respective algorithms for the subprotocols.

\subsection{Extractable range proof for the challenge}\label{app:rangeproof}

During the key generation of $\oschiii$, the server has to prove to the client that the encrypted challenge $B=\mathcal{E}_\ek(\beta)$ has been correctly generated, meaning that $0\leq \beta < 2^\rho$. This proof, together with the necessary trapdoors can be constructed from standard building blocks. The proof requires a hash function $H_\mathsf{com}$, mapping into the group $\GG$, and modelled as a random oracle.

Let $h=H_\mathsf{com}(B,\mathit{ctx})$, where $\mathit{ctx}$ binds this range proof to the current key generation session. E.g. $\mathit{ctx}$ may contain the public key $\pk$ that has already been constructed at that time. Using $h$ as the public key in ElGamal encryption scheme, the server encrypts (in the exponent) each bit of $\beta$ separately. I.e. if $\sum_{i=0}^{\rho-1}\beta_i\cdot 2^i=\beta$, then the server picks $r_0,\ldots,r_{\rho-1}\sample\ZZ_p$ and computes $c_{0,i}=g^{r_i}$, $c_{1,i}=g^{\beta_i}h^{r_i}$ for each $i$. The server sends all values $c_{0,0},\ldots,c_{1,\rho-1}$ to the client. Both the server and the client define $C=\prod_{i=0}^{\rho-1} c_{1,i}^{2^i}$. In this way, $C$ will be a Pedersen commitment to $\beta$, where the randomness is $z=\sum_{i=0}^{\rho-1}r_i\cdot 2^i$. Server proves to the client that the Paillier ciphertext $B$ and Pedersen commitment $C$ contain the same plaintext.

For each bit $\beta_i$, the server proves to the client that it, i.e. the plaintext for the ciphertext $(c_{0,i},c_{1,i})$ is indeed a bit. Note that if $\beta_i=0$, then $(g,h,c_{i,0},c_{i,1})$ is a Diffie-Hellman tuple and the server could convince the client in this using the $\Sigma$-protocol for DDH proofs (Sec.~\ref{sec:preliminaries}); the server knows the randomness $r_i$ necessary to give the proof. Similarly, if $\beta_i=1$, then $(g,h,c_{i,0},c_{i,1})/g$ is a Diffie-Hellman tuple and the same protocol could be used again. In order to show that $\beta_i$ is either $0$ or $1$, we combine these two $\Sigma$-protocols using the construction in App.\ref{app:sigmadisjunct}.

The simulator can decrypt the ElGamal ciphertexts $(c_{i,0},c_{i,1})$ and find the bits $\beta_i$, because it can program $H_\mathsf{com}$. Whenever $H_\mathsf{com}$ is invoked either by itself or by the adversary, the simulator will generate the answer as $g^r$, where $r\sample\ZZ_p$, and store $r$ together with the argument(s) of $H_\mathsf{com}$.

\section{Proving that a DVNIZK proof is valid}\label{app:validdvnizk}

Given a generator $g$ of $\GG$, a Paillier encryption key $\ek$, a ciphertext $B$, and the values $(u,\alpha_1,\Gamma)$, suppose that the prover wants to convince the verifier that they have been constructed as in $\oschiii.\enca$ and the prover knows the used randomness $r=\log_g u\in\ints{d}$, $r_1=\log_g \alpha_1\in\ints{\rho+d+\kappa}$, and the random coins $\mathbf{r}$ for the encryption. The $\Sigma$-protocol for this is similar to proving the knowledge of exponent, where the prover constructs another similar triple $(\alpha_2,\alpha_3,A)$ using fresh randomness $(r_2,r_3,\mathbf{r}')$ and sends it to the verifier, also sends a linear combination (with coefficients chosen by the verifier) of the original and fresh randomness to the verifier, who can then use the homomorphic properties of the operations to match the original triple against the fresh one. \alisachanges{Instead of including $\alpha_2$, $\alpha_3$ and $A$ in the proof, the prover instead includes $\beta'$ and lets the verifier compute $\alpha_2, \alpha_3$ and $A$ that would satisfy the equalities, followed by a check if $\beta'$ indeed computed as their hash.} We present the protocol in Fig.~\ref{fig:validdvnizk} as a NIZK proof, replacing verifier's choice of the coefficients of the linear combination with the invocation of a random oracle. The simulator is thus able to fake these proofs.

\begin{figure}
\begin{pchstack}
\begin{pcvstack}
\pseudocode[head=$\dvp{r}{r_1}{\mathbf{r}}{g}{u}{\alpha_1}{B}{\Gamma}$]{
r_2\sample\ints{\rho+d+\kappa}\\
r_3\sample\ints{2\rho+d+2\kappa}\\
\mathbf{r}'\sample\coins\\
\alpha_2\gets g^{r_2\bmod p}\\
\alpha_3\gets g^{r_3\bmod p}\\
A\gets \mathcal{E}_\ek(r_3;\mathbf{r}')\cdot B^{r_2}\\
\beta'\gets H(g,u,\alpha_1,\Gamma,\alpha_2,\alpha_3,A,B)\\
\gamma_2\gets r_2+\beta'\cdot r\\
\gamma_3\gets r_3+\beta'\cdot r_1\\
\gamma_\mathbf{c}\gets \mathbf{r}'\boxplus (\beta' \boxdot \mathbf{r})\\
\pi\gets(\alisachanges{\beta'},\gamma_2,\gamma_3,\gamma_\mathbf{c})\\
\pcreturn \pi
}
\end{pcvstack} \pchspace[-2em]
\begin{pcvstack}
\pseudocode[head=$\dvc{\pi}{g}{u}{\alpha_1}{B}{\Gamma}$]{
(\alisachanges{\beta'}\comma{} \gamma_2\comma{} \gamma_3\comma{} \gamma_\mathbf{c}) \gets\pi\\
\alisachanges{\alpha_2\gets g^{\gamma_2} / u^{\beta'}}\\
\alisachanges{\alpha_3\gets g^{\gamma_3} / \alpha_1^{\beta'}}\\
\peeterchanges{A\gets \mathcal{E}_\ek(\gamma_3;\gamma_\mathbf{c})\cdot B^{\gamma_2} / \Gamma^{\beta'}}\\
\alisachanges{\pcassert \beta'=H(g,u,\alpha_1,\Gamma,}\\
\peeterchanges{\hspace*{22mm}\alpha_2,\alpha_3,A,B)}
}
\end{pcvstack}
\end{pchstack}
\caption{Proving the validity of a DVNIZK proof}\label{fig:validdvnizk}
\end{figure}

The proof uses the fact that Paillier's encryption is homomorphic with respect to the used coins. In Fig.~\ref{fig:validdvnizk}, the operations $\boxdot$ and $\boxplus$ combine the coins. The value $\mathbf{r}$ is perfectly masked by $\mathbf{r}'$. The values $r,r_1$ however, are masked not perfectly, but only statistically, with random values having $\rho+\kappa$ more bits.

\alisachanges{We show that the protocols of Fig.~\ref{fig:validdvnizk} prevent selective failure attacks on the underlying DFN proofs from~\cite{cryptoeprint:2017/1029}. In a DFN proof, the server verifies a linear combination $\gamma_1 = r_1 + r\beta$. The proof fails if $r_1 + r\beta \geq N$ where $N$ is the size of the plaintext space of the homomorphic encryption scheme. The prover (in our case, the encrypting party) may take $r_1 = N - r\cdot b$ to check whether $\beta \geq b$ by observing whether the verifier (in our case, the server) rejects the proof. The verification procedure of Fig.~\ref{fig:validdvnizk} only depends on $\beta$ through $B =\mathcal{E}_\ek(\beta)$ which does not leak $\beta$ by properties of the homomorphic encryption scheme. However, we need to ensure that Fig.~\ref{fig:validdvnizk} does not miss selective failure attacks, and we show that either $\gamma_2$ or $\gamma_3$ will be too large so that they will rejected by the client. Let $\gamma_1 = r_1 + \beta r \geq N$. This means that either $r_1 \geq N/2$ or $\beta r \geq N/2$. In the first case, we have $\gamma_3 = r_3 + \beta' r_1 \geq \beta' r_1 \geq \beta' N/2$, which gives $\gamma_3 < N$ with probability at most $2^{-\rho}$ (since $\beta'$ is not under attacker's control). In the second case, we have $\gamma_2 = r_2 + \beta' r \geq \beta' r \geq \beta'/\beta \cdot  N/2 \geq \beta' \cdot N / 2^{\rho+1}$. While such $\gamma_2$ can be smaller than $N$, it will still quite likely be larger than the largest possible $\gamma_2$ that can be obtained for a valid proof. A valid $\gamma_2$ by construction is strictly smaller than $2^{\rho + d + \kappa + 1}$, while $N / 2^{\rho+1} \geq 2^{2\rho + d + 2\kappa} / 2^{\rho + 1} \geq 2^{\rho + d + \kappa}$, so the malicious $\gamma_2$ will be smaller than the allowed upper bound with probability $2^{-\rho}$ (again, since $\beta'$ is not under attacker's control).
}

\section{Full proof of Theorem~\ref{thm:main}}\label{app:fullproof}

In order prove security of our threshold decryption protocol in the UC model, we construct a simulator $\simm$ such that for each adversary $\adv$ attacking the protocol, the environment $\zenv$ cannot distinguish whether it is interacting with the real protocol and the adversary $\adv$, or the ideal functionality $\fideal$ and the ``ideal'' adversary (consisting of $\adv$ and $\simm$). Details of $\simm$ are presented in this section.

\lizachanges{We start with defining $\simm$ and how it interacts with the ideal functionality and the adversary in case of 
different parties being corrupted. $\simm$ receives an one-more CDH challenge as input. We proceed with arguing that encryption and decryption performed by the ideal functionality interacting with $\simm$ is indistinguishable from the real protocol. If the \alisachanges{environment} is able to distinguish between real protocol and the ideal protocol, \alisachanges{then the simulator has enough information to solve one-more CDH problem.}} 
\alisachanges{Intuitively, we need one-more CDH problem only in the corrupted client case to ensure that the number of decrypted ciphertexts does not exceed the number of backdoor decryption queries allowed by $\fideal$.}

We assume that the corruptions are known in advance, and $\simm$ gets some data from an external challenger. The challenger generates $\sk \sample \ZZ_p$, $\pk \gets g^{\sk}$, and $r_i \sample \ZZ_p$, $u_i \sample g^{r_i}$ for $i \in \set{0,\ldots,n}$ for a parameter $n$.
\begin{itemize}
\item If the client is \emph{corrupted}, $\simm$ receives $\pk$ and $u_0,\ldots,u_n$, such that $(g,\pk,u_0,\ldots,u_n)$ is an instance of a one-more CDH problem. $\simm$ also gets access to an oracle $(\cdot)^{\sk}$.
\item If the client is \emph{honest}, $\simm$ receives $\pk$ and $u_0$, so that $(g,\pk,u_0)$ is an instance of a CDH problem, i.e. one-more CDH problem with $n=0$. $\simm$ takes $u_1=\cdots=u_n \gets u_0$.
\end{itemize}
In addition, in both cases, $\simm$ gets access to the oracle $\chck(\cdot)$.

Throughout the proof, we will for simplicity assume the following:

\begin{itemize}
\item \emph{Programming a random oracle always succeeds}. In reality, programming a random oracle can fail if the value has already been set in the oracle hash table during one of the previous queries by the simulator or adversary. However, this happens with the negligible probability over all the previously made queries to the corresponding oracles. Random oracle programming happens during the simulation of zero-knowledge proofs, where input values, not known to both parties, are random elements of group $\GG$. Programming oracle used for commitments $H_c$ also involves random group element. Finally, there is programming of oracles $H'$ and $H''$ whose input depends on $\pk^r$ for $r \sample \ZZ_p$. Therefore, there is only a negligible probability of input collisions. 
\item \emph{The adversary cannot create fake proofs of knowledge}. The probability of this happening is negligible due to properties of the underlying constructions. Due to special  honest-verifier zero-knowledge, adversary cannot distinguish simulated protocol messages from the real ones. Due to the special soundness, if adversary creates two accepting transcripts for the same witness, simulator will be able to extract it.
\end{itemize}

The simulation of random oracles that are used in the protocol is defined in Figure~\ref{fig:sim-ro}.

\begin{figure}[bp]
\pseudocode[linenumbering, head={Oracle $H_i(x), i \in \{0, \dots , 3,c \}$:}]{
\mbox{If the \peeterchanges{mapping $\{x\mapsto h\}$} is not in hash table $\textsf{T}_i$:} \\
\t \mbox{Generate random $h \sample range(H_i)$} \\
\t \mbox{Add \peeterchanges{$\{x\mapsto h\}$} to $\textsf{T}_i$} \\
\mbox{Return $\textsf{T}_i(x)$} 
} 
\vspace{0.3cm}
\pseudocode[linenumbering, head={Oracle $\Tilde{H}_i(x), i \in \{0, \dots , 3 \}$:}]{
\mbox{If the \peeterchanges{mapping $\{x\mapsto (h,t)\}$} is not in hash table $\Tilde{\textsf{T}}_i$} \\
\t \mbox{Generate random $t \sample \ZZ_p$} \\
\t \mbox{$h \gets g^t$} \\
\t \mbox{Add \peeterchanges{$\{x\mapsto (h,t)\}$} to $\Tilde{\textsf{T}}_i$} \\
\mbox{Return $\peeterchanges{\mathit{fst}(}\Tilde{\textsf{T}}_i(x)\peeterchanges{)}$} 
}
\vspace{0.3cm}\\
The simulation of $H'$ and $H''$ is analogous to $H_i$.
\caption{Simulation of random oracles.}\label{fig:sim-ro}
\end{figure}

\alisachanges{\textbf{Simulation of knowledge extraction}}

In Figure~\ref{fig:exponent}, we have described how knowledge extraction proofs can be used to raise an element $z$ to a power of $r^{\mathsf{d}}$ for $\mathsf{d} \in \{-1,1\}$. Since the proofs are not necessarily generated during the execution of $\fideal$ (e.g. they can be generated in advance), we need to describe separately how $\simm$ acts at the moments when either $\mathcal{A}$ or $\mathcal{Z}$ accesses the corresponding random oracles. We have three instances of knowledge extraction, indexed by $0,1,2$, each using its own random oracles $H_i$ and $\tilde{H_i}$, and $\mathsf{d}_i \in \{-1,1\}$. For each instance, the simulator maintains a table $\knetable{i}$ to store records $(u,v,t)$ such that $v^t = z^{r^{\mathsf{d}}}$ for $z$ of its choice (which depends on $i$) is expected if there is a valid proof of $v = h^{r^{\mathsf{d}}}$. For $i \in \{0,1,2\}$, we define the routine $\fro{i}{}$ as follows:

\underline{At any point of time}, whenever either $\mathcal{A}$ or $\mathcal{Z}$ accesses the oracle $\tilde{H_i}$ with some input $(g,u,ctx)$, if $i \in \{0,1\}$, take $z = \pk$. If $i = 2$, take $z = u' / g^{r'}$ for $(u',\alpha'_1,\Gamma'_1,\Gamma'_2) \gets ctx$ and $r' = (\mathcal{D}_{\vk_1}(\Gamma'_1) - \mathcal{D}_{\vk_2}(\Gamma'_2)) / (\beta_1 - \beta_2)$. Follow instructions from Fig.~\ref{fig:exponent} to prepare $\tilde{H_i}$ for computing $z^{\log_g(u)^{\mathsf{d_i}}}$.

\underline{At any point of time}, whenever either $\mathcal{A}$ or $\mathcal{Z}$ accesses the oracle $H_i$ with an input of the form $(g,h,u,v,\ldots) $that corresponds to a previously called $h = \tilde{H_i}(g,u,ctx)$, take the $t$ that was generated as in Fig.~\ref{fig:exponent} and add $(u,v,t)$ to $\knetable{i}$. At this point, the simulator does not know yet whether $v^t = z^{\log_g(u)^{\mathsf{d_i}}}$ (it depends on the correctness of $v$), but it can be verified after getting the corresponding proof $\pi$. 

When initialized, the simulator creates empty tables $\knetable{i}$ and starts running $\fro{0}{}$, and $\fro{1}{}$. If the client is corrupted, then as soon as the key generation succeeds, $\simm$ also starts running $\fro{2}{}$ (note that, as far as $\simm$ has not generated $\beta_1$, $\beta_2$, $\vk_1$, $\vk_2$, the inputs to $\tilde{H}_2$ can be ignored as generating a valid proof w.r.t. these secret values is not possible). We further describe how the simulator $\simm$ handles and responds to the commands from $\fideal$.

\textbf{Key generation (client and server are both honest)}

If the server and the client are both honest, the adversary may only interrupt the protocol run. The simulator generates all the exchanged messages himself in such a way that the challenge $\pk$ is used as a part of the public key.

\underline{Key generation}.

Upon receiving message $(\textsf{KeyGen},sid)$ from $\fideal$:
    \begin{itemize} 
    \item Simulate (inside $\simm$) messages for both client and server, setting $\pk$ to the value that was input to the simulator, allowing $\adv$ to stop communication.
    \item If $\adv$ stops communication, set $pk = \bot$. 
    \item Otherwise, set $pk = (\pk,\ek,B_1,B_2)$, where $(\ek$, $B_1$, $B_2)$ come from the simulated protocol.
    \item \alisachanges{Send $(\textsf{Key},sid,pk)$ to $\fideal$.} 
    \end{itemize}
    
\textbf{Key generation  (either server or client is corrupted)}

Simulation of key generation is similar for the cases where either client or server is corrupted. Therefore, we present this simulation jointly here. Let $i \in \set{1,2}$ be the index of the honest party (the index of the corrupted party will be $3-i$).

\underline{Key generation}.
\label{sec:key-ken-sim}

Upon receiving message $(\textsf{KeyGen},sid)$ from $\fideal$:
\begin{itemize}
    \item generate a random bit string $com_i \sample range(H_c)$ send it to the adversary as commitment on $\pk_i$ 
    \item once $com_{3-i}$ is received from adversary, search hash table to get $\pk_{3-i}$ \alisachanges{and $\pi_{3-i}$}.
    \item \alisachanges{Verify the proof $\pkchk[H_0,{\tilde H}_0]{\pi_{3-i}}{g}{\pk_{3-i}}{\Box}{-1}$. If it verifies, then $\simm$ takes $(\ldots,v) \gets \pi_{3-i}$ and looks for an entry $(\pk_{3-i}, v, t) \in \knetable{0}$, and takes $\pk_i = v^t = \pk^{\sk_{3-i}^{-1}}$.}

    \item program random oracle $H_c$ such that $H_c(\pk_i) = com_i$
    \item simulate proof of knowledge of $\sk_i$, following the instructions from Figure~\ref{fig:sim-proof-keygen}. 
    \item send proof $\pi_i$ and key share $\pk_i$ to the adversary 
    \item upon receiving $(\pk_{3-i},\pi_{3-i})$, verify the commitment opening $H_c(\pk_{3-i}\alisachanges{\pi_{3-i}}) = com_{3-i}$. If commitment opening does not verify, set $pk = \bot$.
    \item If client is honest proceed as follows:
    \begin{itemize}
        \item upon receiving $(\ek,B_1,B_2)$ and the additional proofs from the adversary, verify them
        \item if proofs verify, extract $\beta$ from proofs, and set $pk = (\pk,\ek,B_1,B_2)$. If proofs do not verify, set $pk = \bot$. 
    \end{itemize}
    \item If server is honest proceed as follows:
    \begin{itemize}
        \item generate \alisachanges{$(\ek_1,\vk_1) \gets \mathcal{K}()$, $(\ek_2,\vk_2) \gets \mathcal{K}()$}
        \item sample $\beta_1,\beta_2\sample\ints{\rho}\text{ s.t. } \beta_1\not=\beta_2$ and compute $B_1\gets\mathcal{E}_{\alisachanges{\ek_1}}(\beta_1), B_2\gets\mathcal{E}_{\alisachanges{\ek_2}}(\beta_2)$
        \item send $(\ek,B_1,B_2)$ with corresponding simulated proofs to the adversary; \alisachanges{take $pk =  (\pk,\ek,B_1,B_2)$.}
    \end{itemize}
    \item \alisachanges{Send $(\textsf{Key},sid,pk)$ to $\fideal$.}
    \item \alisachanges{For $i \in \{1,2\}$, wait for the ouptut $y_i$ of a corrupted party $P_i$ from $\adv$. Send $\foutput{i}{y_i}$ to $\fideal$.}
\end{itemize}

\begin{figure}[ht!]
\begin{pcvstack}
\pseudocode[linenumbering, head={Simulating $\pkexp[H_0,{\tilde H}_0]{\sk_i}{g}{\pk_i}{\Box}{-1}$}]{
\gamma_{\pi}, \beta_{\pi}, t' \sample \ZZ_p \\
\mbox{Program RO $\Tilde{H}$ as $h' = \Tilde{T}(g,pk_i,ctx) = g^{t'}$} \\
v \gets pk_{i}^{t'} \\ 
\alpha_{\pi} := g^{\gamma_{\pi}}/pk_{i}^{\beta_{\pi}}, \alpha'_{\pi} := (v')^{\gamma_{\pi}}/(h')^{\beta_{\pi}} \\
\mbox{Program RO $H$ as $\beta_{\pi} = T(g,h',pk_i,v,\alpha_{\pi},\alpha_{\pi}',ctx)$} \\
\mbox{Return $(\alisachanges{\beta_{\pi}},\gamma_{\pi},v)$}}
\end{pcvstack}
\caption{Simulating proof of knowledge of secret key share}
\label{fig:sim-proof-keygen}
\end{figure}

\begin{figure}[tbp]
\pseudocode[linenumbering, head={Simulating $\pkexp[H_1,{\tilde H}_1]{r}{g}{u}{ctx}{1}$}]{
\gamma_{\pi}, \beta_{\pi}, t' \sample \ZZ_p \\
\mbox{Program RO $\Tilde{H}$ as $h' = \Tilde{T}(g,u,ctx) = g^{t'}$} \\
\mbox{$v \gets u^{t'}$} \\
\alpha_{\pi} \gets g^{\gamma_{\pi}}/u^{\beta_{\pi}}, \alpha'_{\pi} := (h')^{\gamma_{\pi}}/v^{\beta_{\pi}} \\
\mbox{Program RO $H$ as $\beta_{\pi} = T(g,h',u,v,\alpha_{\pi},\alpha_{\pi}',ctx)$} \\
\mbox{Return $(\alisachanges{\beta_{\pi}},\gamma_{\pi},v)$}
}
\vspace{0.3cm}
\pseudocode[linenumbering, head={Simulating $\dvp{r}{r_1}{r_\iota}{g}{u}{\alpha_1}{B_\iota}{\Gamma_\iota}$}]{
\beta'_{\alisachanges{\iota}},\gamma_{(\pi_{\alisachanges{\iota}},2)},\gamma_{(\pi_{\alisachanges{\iota}},3)} \sample \ZZ_p, \gamma_\mathbf{c} \sample \coins \\
\alpha_2 \gets g^{\gamma_{(\pi_{\alisachanges{\iota}},2)}} / u^{\beta'_{\alisachanges{\iota}}}, \alpha_3 \gets g^{\gamma_{(\pi_{\alisachanges{\iota}},3)}} / \alpha_1^{\beta'_{\alisachanges{\iota}}} \\
A \gets (\mathcal{E}_{\alisachanges{\ek_{\iota}}}(\gamma_3; \gamma_\mathbf{c}) \cdot B_{\alisachanges{\iota}}^{\gamma_2})/\Gamma_{\alisachanges{\iota}}^{\beta'_{\alisachanges{\iota}}} \\
\mbox{Program RO $H$ as $\beta'_{\alisachanges{\iota}} = T(g,u,\alpha_1,\Gamma_{\alisachanges{\iota}},\alpha_2,\alpha_3,A,B_{\alisachanges{\iota}})$} \\
\mbox{Return $(\alisachanges{\beta'_{\alisachanges{\iota}}},\gamma_2,\gamma_3,\gamma_\mathbf{c})$}
}
\caption{Simulators for zero-knowledge proofs of encryption}
\label{fig:sim-proof-enc}
\end{figure}

After key generation phase has been simulated, $\simm$ sets a counter $ctr_{\mathsf{chal}} \gets 0$, defining the index of the challenge $u_{ctr_{\mathsf{chal}}}$ that will be used in the next encrypiton.

\textbf{Encryption (any corruptions)}

We assume that the key generation has already been simulated, and $\simm$ has come up with a public key $pk = (\pk,\ek,B_1,B_2)$. The simulator maintains a local table $\mathsf{T}_{\mathsf{i,r,c}}$, storing a record $(i,r,c)$ for a challenge index $i$, the randomness $r$, and the ciphertext $c$ 
for each ciphertext $c$ generated for $\fideal$.

\underline{Encryption}

Upon receiving message $(\textsf{Encrypt}, sid)$ from $\fideal$, compute $c\gets\encasim(j\comma{} \pk\comma{} \ek\comma{} B_1\comma{} B_2)$ for $j = ctr_{\mathsf{chal}}$ as in Fig.~\ref{encryption-function}. Set $ctr_{\mathsf{chal}} \gets ctr_{\mathsf{chal}} + 1$. Send message $(\textsf{Encrypt},sid,c)$ to $\fideal$. 

\alisachanges{Upon receiving message $(\textsf{Encrypt}, sid, i, m)$, simulate input $m$ for $P_i$ in the real protocol and wait for $y$ from $\adv$. Send message $(\textsf{Encrypt},sid,y)$ to $\fideal$.}

\begin{figure}[ht!]
\begin{pcvstack}
\pseudocode[linenumbering, head=$\encasim(i\comma{}\pk\comma{}\ek\comma{}B_1\comma{}B_2)$]{
\gray{r\sample\ZZ_p}\\
u\gets u_i^{r}\text{ where $u_i$ is a challenge}\\
\gray{\mathbf{r}_\iota\sample \coins}\\
\alisachanges{\gamma \sample \ints{\rho + d + \kappa}}\\
\alisachanges{\alpha_1 \gets g^{\gamma} / u^{\beta_1}}\text{if the \lizachanges{client} is honest, sample $\beta_1 \sample \ints{\rho}$}\\
\Gamma_\iota \gets \mathcal{E}_{\alisachanges{\ek_{\iota}}}(\alisachanges{\gamma ; \mathbf{r}_{\iota}}) \\
\text{Simulate }\pi, \pi_{\iota}\text{ as in Fig.~\ref{fig:sim-proof-enc}}\\
\gray{c_1\gets (u,\alpha_1,\Gamma_1,\Gamma_2,\pi,\pi_1,\pi_2)}\\
c_{21} \sample range(H')\\
c_{22} \sample range(H'')\\
c\gets (c_1,(c_{21},c_{22}))\\
\text{Store the record $(i,r,c)$ in a table $\mathsf{T}_{\mathsf{i,r,c}}$}\\
\pcreturn c
}
\end{pcvstack}
\caption{Encryption function $\encasim$ as computed by the simulator. It simulates real protocol encryption $\enca$ (Fig.~\ref{fig:enc-dec}) to get $c_1$. The challenge $u_i$ is embedded into the ciphertext, and all proofs are simulated. The parts of code that are the same as in $\enca$ are colored gray.
The simulator samples random $c_{21}$ and $c_{22}$ so that the RO of $H'$ and $H''$ satisfying $c_{21} = H'(pk^r) \oplus m$ and $c_{22} = H''(pk^r,c_{21})$ could be programmed later. The simulator remembers which challenge and which randomness was used for which ciphertext.
\label{encryption-function}
}
\end{figure}

\textbf{Decryption (honest client)}

\underline{Decryption}.
Upon receiving $(\textsf{Decrypt-init},sid)$ from $\fideal$:
\begin{itemize}
\item Send a message $(\textsf{Decrypt-init}, sid, \ver)$ to $\fideal$, where the algorithm $\ver$ is defined as on Figure~\ref{fig:verify-alg}.
\item Upon receiving $(\textsf{Decrypt-bad-c,sid})$ from $\fideal$, stop the simulation on behalf of the client before any interaction has started. This corresponds to failed proof check. \alisachanges{If the server is corrupted, wait for the server output $y_2$ from $\adv$ and send $\foutput{2}{y_2}$ to $\fideal$.}
\item Upon receiving $(\textsf{Decrypt-good-c,sid})$ from $\fideal$:

If the server is \emph{honest}:
\begin{itemize}
\item Proceed with simulating protocol messages, allowing $\adv$ to interrupt communication.
\item When the protocol simulation finishes, send $(\textsf{Decrypt-complete},sid,\dec)$ to $\fideal$ with $\dec$ defined as in Fig.~\ref{fig:decryption-alg}.
\end{itemize}

If the server is \emph{corrupted}:
\begin{itemize}
    \item Compute $c \gets \enca(\mu,\pk,\ek,B_1,B_2)$ for some fixed messge $\mu \in \plaintxtset$.
    \item \alisachanges{Sample blinding values for $c$ and generate the proof $\pi'$ according to $\decp{C}$ of Fig.~\ref{fig:enc-dec} (note that secret key is not needed for these steps), getting a blinded ciphertext $c'$. Send $c'$ to $\adv$.}
    \item Upon receiving $(w, \pi'')$ from $\adv$ verify $\pi''$.

    If it verifies, send $(\textsf{Decrypt-complete},sid,\dec)$ to $\fideal$ with $\dec$ defined as in Fig.~\ref{fig:decryption-alg}.  
    Otherwise, send $(\textsf{Decrypt-fail},sid)$ to $\fideal$. \alisachanges{Ignore $w$.}
\item \alisachanges{Wait for the server output $y_2$ from $\adv$ and send $\foutput{2}{y_2}$ to $\fideal$.}
\end{itemize}
\end{itemize}

Since the simulator encrypts a fixed message $\mu$ in the simulated protocol, we need that the adversary would not be able to distinguish if a random ciphertext is being decrypted or it corresponds to a blinded version of some ciphertext $c$ that might have been seen previously by the adversary. This refers to privacy property provided for the decrypted ciphertext, and since we have statistical blinding, this probability is negligible.

\textbf{Decryption (corrupted client)}

\underline{Decryption}
Upon receiving $(\textsf{Decrypt-init}, sid, c')$ from $\fideal$:
\begin{itemize}
    \item send $c'$ to the adversary;
    \item upon receiving $(u',\alpha'_1,\Gamma'_1,\Gamma'_2,\pi')$ from the adversary, 
    \alisachanges{if $\pi'$ does not verify, send $(\textsf{Decrypt-fail},sid)$ to $\fideal$};
    \item \alisachanges{If $\pi'$ verifies, compute $\gamma_1\gets \mathcal{D}_{\alisachanges{\vk_1}}(\Gamma'_1)$ and $\gamma_2\gets \mathcal{D}_{\alisachanges{\vk_2}}(\Gamma'_2)$. Extract $r' = (\gamma_1 - \gamma_2) / (\beta_1 - \beta_2)$.} 
    \item \alisachanges{If $g^{r'}=u'$ then compute $w=\pk_2^{r'}$}.
    \item \alisachanges{Otherwise, compute $u'' = u' / g^{r'}$ and $\gamma' = \gamma_1 - \beta_1 r'$ (note that $\gamma' = \gamma_2 - \beta_2 r'$ due to choice of $r'$). Verify $\gamma' = \alpha'_1 \cdot (u'')^{\beta_1}$. If this check fails, send $(\textsf{Decrypt-fail},sid)$ to $\fideal$. If this check passes, then the client has come up with $\Gamma'_2 = \mathcal{E}_{\ek_2}(\log_g(\alpha'_1) + \log_g(u'') \cdot \beta_1 + \log_g(u'/u'')\cdot \beta_2)$ for the provided $\alpha'_1$ and $u'$. Unless $u'' = 1$, the client needs quantities of the form $\mathcal{E}_{\ek_2}(r_j \cdot \beta_1)$, which can only be taken from challenge ciphertexts where $r_j = \log_g(u_j)$ for a challenge $u_j$, so $u'' = \prod_j u_j^{z_j}$ for such $u_j$ and some linear coefficients $z_j$.}
    
    \item \alisachanges{As the proof $\pi' = (\pi,v)$ has verified}, there is \alisachanges{$(\pk_1,v,t) \in \knetable{2}$ such that $v^t = u' / g^{r'} = u''$};
    \item \alisachanges{Call the oracle $(\cdot)^{\sk}$ at the point $w' := v^t$, getting $w'' = ((u'')^{1/sk_1})^{\sk} = (u'')^{\sk_2}$. Take $w = w'' \cdot \pk_2^{r'}$. At this point, the oracle $(\cdot)^{\sk}$ has only been used on linear combinations (with coefficients known to the adversary) on challenge ciphertexts, so the result cannot be used to break any assumptions, e.g. fake some proofs.} 
    \item simulate proof $\pi''$ following the instructions of Figure~\ref{fig:sim-proof-dec} and send $(w,\pi'')$ to the adversary;
    \item Send $(\textsf{Decrypt-complete},sid)$ to $\fideal$. 
    \item \alisachanges{Upon receiving the final output $y_1$ from $\adv$, send $\foutput{1}{y_1}$ to $\fideal$.}
    \end{itemize}    
    
\textbf{Offline decryption}

Before proving correctness of the simulations for encryption and decryption, we need to describe how the simulator resolves adversary's attempts to decrypt a ciphertext $c$ offline, i.e. without involving $\fideal$. If the client is corrupted, then in addition to $\mathsf{T}_{i,r,c}$ the simulator maintains a local table $\mathsf{T}_{\mathsf{m,c}}$ storing records $(m,c)$ for message-ciphertext pairs it already knows. 

\underline{At any point of time} whenever either $\adv$ or $\zenv$ queries $H'$ at the point $x$ or $H''$ at the point $(x,y)$ for any $y$, if $H'(x)$ has not been programmed yet:
        \begin{itemize}
            \item Iterate over records $(i,r_i,c_i)$ of the table $\mathsf{T}_{\mathsf{i,r,c}}$.
            \item For each $r_i$, take $x' \gets x^{1/r_i}$, and call $\chck(x')$. This will tell whether $x$ is a solution to some component of the one-more CDH problem, i.e. corresponds to a ciphertext generated by $\simm$ for $\fideal$.
            \item If the check succeeds for some $i$, returning $j$, then $x$ is a solution to the CDH problem $(g,\pk,u_j)$. 
                  \begin{itemize}
                  \item If the client is \emph{honest}, then $u_j = u_0$ for all $j$, and the simulator has solved the CDH problem $(g,\pk,u_0)$.
                  \item If the client is \emph{corrupted}, then $\simm$ may not have solved one-more CDH problem yet. If this is the case, $\simm$ proceeds as follows:
                      \begin{itemize}
                      \item let $(c_{i1}, (c_{i2},c_{i3})) \gets c_i$;
                      \item if $\mathsf{T}_{\mathsf{m,c}}$ does not contain a record $(m,c_i)$, then send $(\textsf{Decrypt-msg},sid,\bot,c_i)$ to $\fideal$, wait for response $(\textsf{Decrypted},sid,m)$, and add $(m,c_i)$ to $\mathsf{T}_{\mathsf{m,c}}$ ;

                           \item program random oracle $H'$ s.t. $H'(x) = m \oplus c_{i2}$;
                           \item program random oracle $H''$ s.t. $H''(x,c_{i2}) = c_{i3}$.

                      \end{itemize}
                  \end{itemize}
        \end{itemize}

\begin{figure}[ht!]
\begin{pcvstack}
\pseudocode[linenumbering, head={Simulating $\dhp[H_3]{\sk_2}{\pk_1}{u'}{\pk}{w}{\Box}$}]{
\gamma_{\pi''}, \beta_{\pi''} \sample \ZZ_p \\
\alpha_{\pi''} \gets pk_1^{\gamma_{\pi''}}/pk^{\beta_{\pi''}}, \alpha'_{\pi''} \gets (u')^{\gamma_{\pi''}}/(w')^{\beta_{\pi''}} \\
\mbox{Program RO $H$ $\beta_{\pi''} = T(pk_2,pk,w,u',\alpha_{\pi''},\alpha_{\pi''}')$} \\
\mbox{Return $(\alisachanges{\beta_{\pi''}},\gamma_{\pi''})$}}
\end{pcvstack}
\caption{Simulating proof of knowledge of secret key share for server decryption}
\label{fig:sim-proof-dec}
\end{figure}

\textbf{Proofs of indistinguishability}

We prove that encryption and decryption performed by $\fideal$ output to the receiving party the values that would be expected in a real protocol.

\underline{Encryption}. The ciphertext $c' \gets \encasim(i,\pk,\ek,B_1,B_2)$ output by $\fideal$ is indistinguishable from a real ciphertext $c \gets \enca(m,\pk,\ek,B_1,B_2)$ as far as $H'(u_i^{r\cdot\sk}) \oplus m$ is indistinguishable from a random element of $range(H')$, and $H''(u_i^{r\cdot\sk},c_2)$ from a random element of $range(H'')$. This distinguishing is possible in the following cases:
        \begin{itemize}
        \item The hash function $H'$ (or $H''$) does not return a random element of $range(H')$ (or $range(H'')$). We assume that $H'$ and $H''$ act as random oracles.
        \item The adversary could obtain $u_i^{r\cdot\sk}$. In this case, $\simm$ would receive $u_i^{r\cdot\sk}$ at the point where the adversary queries $H'(x)$ or $H''(x,y)$ for $x = u_i^{r\cdot\sk}$. Acting according to the instructions for the offline decryption, he would either solve one instance of the one-more CDH problem, or program $H'$ and $H''$ so that $c$ decrypts to $m$.
       \end{itemize}
       If the client is \emph{honest}, then solving one instance of one-more CDH problem is enough to answer the challenge. If the client is \emph{corrupted}, since $\simm$ is allowed to get at most $\mathit{ctr}_\mathsf{dec}$ messages $m$ from $\fideal$, we need to show that $\mathit{ctr}_\mathsf{dec}$ is sufficiently large. 

        W.l.o.g. suppose that the adversary manages to obtain $u_1^{r_1\cdot\sk},\ldots,u_\ell^{r_{\ell}\cdot\sk}$. We show that $\ell \leq n_d$ where $n_d$ is the number of decryption sessions initiated so far. In each decryption session, $\simm$ accesses the oracle $(\cdot)^\sk$ at most once, and there are no other accesses to $(\cdot)^{\sk}$, so $n = n_d$. For $\ell$ different challenges $u_i^{r_i\cdot \sk}$, there will be needed $\ell$ queries to $\fideal$ to get $m$. If $\ell > n = n_d$, then $\simm$ has solved one-more CDH problem of size $n_d$. If $\ell \leq n = n_d$, then $\mathit{ctr}_\mathsf{dec}$ stays non-negative.

\underline{Decryption}. The message $m$ resulting from decryption of $c$ by $\fideal$ should be indistinguishable from what would be expected in the real protocol. This is relevant only for the honest client case, as the output of a corrupted client is determined by the adversary. There are different cases for $c$:

     \begin{itemize}
        \item If $c$ was output to the environment by $\fideal$ as a response to an encryption query, then $\fideal$ has a valid pair $(m,c)$ in the table \textsf{T} and does not need to apply $\dec$ algorithm.
        \item If $c$ was generated externally, then it is being decrypted using algorithm $\dec$ which \alisachanges{computes $\pk^r$ as $v^t$ for an entry $(u,v,t)$ of the table $\knetable{1}$ for $u$ and $\pi$ that are a part of $c$. This record is always there since $\dec$ is applied only in the cases where $\ver$ has been successful, which means that the corresponding proofs have been generated and oracles $H_1$, $\tilde{H}_1$ accessed.} 
        If $c$ depends on some ciphertext $c'$ generated for a message $m'$ using $\fideal$, the adversary could notice that $c'$ does not depend on $m'$. We show that $c$ cannot be related to $c'$ unless either $c = c'$, or the only part that differs from $c$ are the components of $c_1$ that are different from $u$, which means that $\dec$ would decrypt $c$ using \alisachanges{$\mathsf{T}$}.
        \end{itemize}
        
        Suppose that the adversary is able to get a ciphertext $c = (c_1,(c_2,c_3))$ acceptable for decryption by $\fideal$. 
        Let $c' = (c'_1,(c'_2,c'_3))$ be any ciphertext generated using $\fideal$.

        \begin{enumerate}
                
        \item Suppose the adversary has constructed $\pi$ itself.
        The other components of $c$ may depend on $c'$. 
However, since $\pk^r$ can be extracted from $\pi$, the ciphertext $c$ could be decrypted \emph{without} using $\fideal$, and we proved in the previous point indistinguishability for $c'$ in this case. 

        \item Suppose the adversary has taken the proof $\pi$ from $c'$. Since $\pi$ has been generated w.r.t. $u_i^{r_i}$, $\alpha_1,\Gamma_{\iota}$, the adversary would need to keep these components the same. 
        Due to the assertion $c'_3 = H''(u_i^{r_i\cdot \sk},c'_2)$ that an honest client would perform in the real protocol before outputting $m$, the adversary needs to compute $H''(u_i^{r_i\cdot \sk}, c_2)$ to take $c_2 \neq c'_2$. This would require access to $H''$ with $u_i^{r_i\cdot \sk}$, which would allow $\simm$ to solve CDH. If $c_1 = c'_1$ and $c_2 = c'_2$, then also $c_3 = c'_3$.
        \qedhere
        \end{enumerate}

\end{document}